\documentclass[journal]{IEEEtran}

\ifCLASSINFOpdf

\else

\fi
\usepackage{soul}
\usepackage[utf8]{inputenc}
\usepackage{amsmath}
\usepackage{mathtools}
\usepackage{amsthm}

\usepackage{braket}
\usepackage{graphicx}
\usepackage{tikz}
\usepackage{quantikz}
\graphicspath{{./figures/}}
\usepackage{amsfonts}
\usepackage{comment}
\usepackage[ruled,vlined,linesnumbered]{algorithm2e}
\let\oldnl\nl
\newcommand{\nonl}{\renewcommand{\nl}{\let\nl\oldnl}}

\usepackage{csquotes}
\usepackage{hhline}
\usepackage{amssymb}
\usepackage{listings}
\usepackage{color}
\definecolor{codegreen}{rgb}{0,0.6,0}
\definecolor{codegray}{rgb}{0.5,0.5,0.5}
\definecolor{codepurple}{rgb}{0.58,0,0.82}
\definecolor{backcolour}{rgb}{0.95,0.95,0.92}

\usepackage{subcaption}
\usepackage{titlesec}

\usepackage{dcolumn}
\usepackage{tabularx}
\usepackage{hyperref}
\hypersetup{
    colorlinks=true,
    linktoc=all,
    linkcolor=black,
    citecolor=magenta,
}
\usepackage{longtable}
\usepackage{braket}
\newcolumntype{C}{>{\centering\arraybackslash}X}
\SetAlFnt{\small}
\usepackage[font=small]{caption}
\begin{document}
\hyphenation{op-tical net-works semi-conduc-tor}

\title{Robust Pretty Good Measurement via Hybrid Classical-Quantum Pseudoinverse Approximation and Circuit-Level Realization}

\author{Bikash~K.~Behera\thanks{B.~K. Behera is with Università degli Studi di Cagliari, Via Is Mirrions, Cagliari, 09123, Italy., e-mail: (bikas.riki@gmail.com).}, Andrés Camilo Granda Arango\thanks{A.~C.~G. Arango is with Università degli Studi di Cagliari, Via Is Mirrions, Cagliari, 09123, Italy, e-mail: (andresgran34@gmail.com).}, Giuseppe Sergioli\thanks{G. Sergioli is with Università degli Studi di Cagliari, Via Is Mirrions, Cagliari, 09123, Italy, e-mail: (giuseppe.sergioli@gmail.com).}, and Roberto Giuntini\thanks{R. Giuntini is with Università degli Studi di Cagliari, Via Is Mirrions, Cagliari, 09123, Italy, and Technische Universität München. Institute for Advanced Study (IAS), Lichtenbergstraße 2a, 85748 Garching b. M\"unchen, Germany, e-mail: (giuntini@unica.it).}}%

\maketitle

\begin{abstract}
Pretty Good Measurement (PGM) is a near-optimal strategy for quantum state discrimination, but its practical realization becomes unstable when the ensemble operator is singular or ill-conditioned, situations commonly encountered in realistic datasets and noisy quantum systems. In this work, we introduce a numerically consistent formulation of PGM based on the Moore–Penrose pseudoinverse, replacing the standard inverse square root with a threshold-regularized variant to ensure numerical stability across all spectral regimes. Building on this formulation, we develop a hybrid classical–quantum framework that combines pseudoinverse-based spectral preprocessing with constructive quantum circuit realizations based on block-encoding and spectral transformation techniques. The framework incorporates support awareness, restricting operations to the effective subspace of the ensemble operator and yielding well-defined measurement operators even for rank-deficient cases. To further enhance practical feasibility, we integrate an oblivious amplitude amplification step that significantly improves success probabilities in the circuit-level implementation. The proposed approach realizes the pseudoinverse-induced spectral transformation through executable quantum circuits while maintaining the pseudoinverse construction as a classical preprocessing stage. We validate the framework through extensive numerical and quantum circuit simulations, demonstrating close agreement between theoretical predictions and circuit outcomes, with total variation distances as low as $10^{-6}$ in representative experiments. Experiments on both synthetic and real datasets, including ill-conditioned and degenerate scenarios, show that the method achieves stable and reliable discrimination performance where standard PGM becomes numerically unstable. These results establish a practical hybrid classical–quantum framework for robust quantum state discrimination, extending previous circuit-based implementations of the PGM testing stage toward pseudoinverse-aware measurement design and providing a pathway toward scalable quantum-enhanced measurement protocols on noisy intermediate-scale quantum devices.
\end{abstract}


\IEEEpeerreviewmaketitle

\section{Introduction}

Quantum computing has emerged as a transformative paradigm capable of addressing computational challenges that are intractable for classical systems. By leveraging quantum mechanical phenomena such as superposition, entanglement, and interference, quantum algorithms have demonstrated the potential to achieve exponential or polynomial speedups over their classical counterparts in domains such as optimization, cryptography, and machine learning \cite{nielsen2010quantum, montanaro2016quantum}. One of the central problems in quantum information theory is \textit{quantum state discrimination}, which aims to distinguish between non-orthogonal quantum states with maximum accuracy \cite{chefles2000quantum, barnett2009quantum}. Among the various measurement strategies, the Pretty Good Measurement (PGM), also known as the square-root measurement, has received significant attention due to its near-optimal performance and analytical tractability \cite{hausladen1994pretty, barnum2002reversing}. PGM has been widely studied in contexts such as quantum communication, channel decoding, and quantum machine learning, where efficient and reliable discrimination of quantum states is essential. More broadly, quantum state discrimination plays a fundamental role in several areas of quantum information processing, including quantum computational logics and the study of non-classical information structures, where quantum states and measurements provide the semantic basis for logical inference and decision processes \cite{dallachiara2018manyvalued}. Despite its theoretical appeal, the practical implementation of PGM remains challenging, particularly when the ensemble operator becomes singular or ill-conditioned. In such regimes, the standard formulation based on matrix inversion becomes unstable or undefined, limiting its applicability in realistic scenarios involving noise, degeneracy, or rank-deficient data. These conditions naturally arise in real-world datasets with strong correlations or limited diversity, as well as in Noisy Intermediate-Scale Quantum (NISQ) devices where imperfections and spectral degeneracies are unavoidable. Consequently, standard PGM constructions may exhibit numerical instability and degraded performance, hindering their use in practical quantum algorithms.

Several works have explored robust quantum measurement strategies and generalized formulations of state discrimination to address these issues \cite{eldar2003mixed, bae2015quantum}. Previous work by Giuntini et al. \cite{Giuntini2023QuantumClassification} introduced a quantum-inspired multi-class classifier based on PGM, where the measurement operators were constructed from the training dataset and subsequently evaluated through quantum-circuit implementations of the testing stage. While that work demonstrated the feasibility of PGM-based classification and circuit-level evaluation, the construction of inverse-type operators associated with the ensemble state remained outside the quantum-circuit framework. The present work extends this direction by developing a support-aware pseudoinverse-based formulation together with a hybrid classical–quantum implementation strategy for the corresponding spectral transformations. In parallel, quantum linear algebra techniques have been extensively investigated as a foundation for quantum algorithms. The Harrow-Hassidim-Lloyd (HHL) algorithm provides a theoretical exponential speedup for solving linear systems under certain assumptions \cite{harrow2009quantum}. Subsequent works have focused on improving its practicality by addressing key bottlenecks such as state preparation, Hamiltonian simulation, and eigenvalue inversion \cite{childs2017quantum, berry2015hamiltonian}. More recently, the development of block-encoding and quantum singular value transformation (QSVT) has provided a powerful framework for implementing matrix functions on quantum computers with provable guarantees \cite{gilyen2019quantum, low2017optimal}. These advances offer a promising pathway for realizing complex operator transformations required in quantum measurement design. However, the practical deployment of such techniques remains constrained by noise, limited qubit counts, and circuit depth. Hybrid quantum-classical approaches and efficient circuit design methodologies have therefore gained significant attention \cite{cerezo2021variational, schuld2015introduction}. In this direction, quantum-inspired approaches based on quantum state discrimination have been successfully applied to classification tasks. In particular, Sergioli \textit{et al.} \cite{sergioli2019} introduced a quantum framework for binary classification using density operator representations. These ideas were further extended to multi-class scenarios by Giuntini \textit{et al.} \cite{Giuntini2023QuantumClassification, giuntini2023multiclass_asoc}, where quantum-inspired algorithms were developed for direct classification based on state discrimination principles. While these approaches demonstrate the effectiveness of measurement-based classification, they also highlight the need for robust and scalable implementations, especially in the presence of ill-conditioned or noisy quantum representations. For instance, Dutta \textit{et al.} \cite{Dutta2020Regression} proposed a systematic framework for quantum circuit design in regression tasks, emphasizing the importance of efficient encoding and structural optimization.

Motivated by these developments, this work focuses on developing a robust and circuit-compatible formulation of PGM for quantum state discrimination under ill-conditioned ensembles. We address key limitations of existing approaches by introducing a support-aware and threshold-regularized pseudoinverse framework, ensuring operational validity and spectral consistency even in rank-deficient regimes. Building on recent advances in quantum linear algebra, we further develop a hybrid classical-quantum realization based on block-encoding and spectral transformation techniques, where the spectral preprocessing and phase-generation stages are performed classically and the resulting transformations are implemented through executable quantum circuits. To enhance practical feasibility, we incorporate an amplitude amplification step that improves success probabilities in the circuit-level realization. Unlike prior works, this study provides both a mathematically robust formulation and a hybrid classical-quantum implementation framework for pseudoinverse-based PGM construction. The proposed framework is validated through numerical and circuit-level simulations, demonstrating very close agreement between theoretical predictions and quantum circuit outputs, even in ill-conditioned regimes. This establishes a direct bridge between abstract measurement theory and executable quantum circuits, enabling reliable quantum state discrimination in realistic settings.

The main contributions of this work can be summarized as follows:
\begin{itemize}

\item We develop a support-aware and threshold-regularized formulation of PGM based on the Moore-Penrose pseudoinverse, enabling stable and well-defined operation for ill-conditioned and rank-deficient quantum ensembles.

\item We provide a physically consistent operator-level characterization of the proposed construction, including a Kraus-operator representation that serves as a physical consistency check and supports circuit-level realization.

\item We design a hybrid classical-quantum framework that combines classical spectral preprocessing, pseudoinverse regularization, and phase optimization with circuit-level realizations based on block-encoding and spectral-transformation techniques.

\item We perform end-to-end validation through quantum circuit simulations on both real and synthetic datasets, demonstrating near-perfect agreement between theoretical predictions and circuit outputs.

\item We systematically analyze robustness under ill-conditioning, noise, threshold regularization, and finite-shot effects, revealing key trade-offs between discrimination performance, numerical stability, and implementation complexity.

\end{itemize}

The remainder of this paper is organized as follows. Section~\ref{background} reviews the theoretical foundations of quantum state discrimination and PGM, including the role of the Moore-Penrose pseudoinverse in singular and ill-conditioned regimes. Section~\ref{methodology} introduces the proposed support-aware and threshold-regularized PGM framework, together with its circuit-oriented realization based on block-encoding and an equivalent Kraus-operator representation. Section~\ref{results} presents extensive experimental results, including both matrix-level analysis and circuit-based simulations on real and synthetic datasets, along with detailed studies of robustness under noise, threshold regularization, and finite-shot effects. Finally, Section~\ref{conclusion} concludes the paper and outlines directions for future research.

\section{Background}\label{background}

The PGM provides an optimal or near-optimal strategy for quantum state discrimination and has been extensively studied since its original formulation by Hausladen and Wootters~\cite{hausladen1994pretty}. In particular, PGM is known to achieve optimal performance for several classes of symmetric ensembles and provides strong approximation guarantees in general discrimination tasks. This makes it a central tool in quantum information theory, with applications ranging from quantum communication and hypothesis testing to quantum machine learning and classification. Despite its theoretical importance, the practical realization of PGM in a quantum circuit model remains highly nontrivial. While recent works have demonstrated efficient classical evaluation and partial quantum testing of PGM-based classifiers, a fully constructive implementation of the PGM protocol at the circuit level is still lacking. The primary technical challenge arises from the need to implement matrix inverse or inverse square-root operations associated with the ensemble operator. These transformations are inherently non-unitary and require careful spectral manipulation. While recent advances in block-encoding and QSVT provide a framework for implementing such spectral functions, practical realizations typically require a combination of classical spectral preprocessing and quantum circuit execution.

This challenge naturally connects to quantum linear systems algorithms (QLSA), most notably the HHL algorithm~\cite{harrow2009quantum}, which provides a framework for implementing matrix inversion via phase estimation and controlled rotations. Subsequent developments in quantum linear algebra have introduced more flexible and efficient approaches for implementing spectral transformations, including block-encoding and QSVT, which enable polynomial approximations of matrix functions within a unitary framework~\cite{gilyen2019quantum, low2017optimal}. These techniques provide a powerful foundation for realizing nontrivial operator functions such as inverse square roots and pseudoinverse-related spectral transformations, which are essential for PGM and related state-discrimination protocols. A further complication arises in realistic settings, where quantum datasets often lead to ensemble operators that are not perfectly conditioned. In such cases, naive inversion becomes unstable or even undefined, particularly in the presence of small or vanishing eigenvalues. This motivates the need for a more robust formulation of PGM that remains well-defined across all spectral regimes. In this work, we address this limitation by adopting the Moore–Penrose pseudoinverse as a mathematically consistent and operationally meaningful alternative to standard inversion.

The closest related work is the quantum-inspired PGM classifier introduced in \cite{Giuntini2023QuantumClassification}. There, the training phase was performed through classical construction of density operators and PGM measurement elements, while quantum circuits were employed to evaluate the resulting classification functions. In contrast, the present work focuses on pseudoinverse-based transformations associated with the ensemble operator and develops a hybrid classical-quantum framework in which spectral preprocessing is performed classically and the resulting transformations are represented and validated through circuit-level realizations based on block-encoding and spectral-transformation techniques. Therefore, the current work should be viewed as an extension of the circuit-based testing framework introduced in \cite{Giuntini2023QuantumClassification}, providing a hybrid classical-quantum treatment of pseudoinverse-based ensemble transformations rather than a fully quantum realization of the entire PGM training process.

\subsection{PGM Formalism, Support Restriction, and the Role of the Pseudoinverse}

Given an ensemble $\{p_i, \rho_i\}_{i=1}^K$ of quantum states $\rho_i$ with prior probabilities $p_i$, the average state (also called the ensemble operator) is defined as
\begin{equation}
S = \sum_{i=1}^K p_i \rho_i.
\label{Eq1}
\end{equation}
The operator $S$ is positive semidefinite and encodes the global structure of the ensemble. The PGM POVM elements are defined as
\begin{equation}
M_i = S^{-1/2} (p_i \rho_i) S^{-1/2},
\end{equation}
which satisfy the completeness relation $\sum_i M_i = I$ when $S$ is full rank. The effectiveness of PGM arises from its ability to approximate the Helstrom optimal measurement while maintaining a relatively simple analytical structure. In realistic scenarios, however, the operator $S$ may be rank-deficient or ill-conditioned. This occurs, for example, when the ensemble states span only a subspace of the full Hilbert space or when there is significant overlap between states. In such cases, the inverse $S^{-1}$ is either undefined or highly unstable, leading to unreliable measurement operators. To address this issue, the Moore–Penrose pseudoinverse must be used instead of the ordinary inverse, yielding
\begin{equation}
M_i = S^{+1/2} (p_i \rho_i) S^{+1/2}.
\end{equation}
Here, $S^{+1/2}$ denotes the square root of the pseudoinverse of $S$, which acts as a well-defined inverse on the support of $S$ and annihilates its null space. In practice, the pseudoinverse acts only on the support of $S$, defined by the projector
\begin{equation}
\Pi_S=\sum_{\lambda_j>0}\ket{u_j}\bra{u_j},
\end{equation}
where $\{\lambda_j,\ket{u_j}\}$ are the eigenpairs of $S$. Consequently, all inverse-type operations are restricted to the support subspace, while components belonging to the null space are removed. This support-aware interpretation is essential for obtaining stable and physically meaningful PGM operators in rank-deficient regimes. The pseudoinverse was independently introduced by Moore~\cite{Moore1920GeneralInverse} and Penrose~\cite{Penrose1955PseudoInverse}, and it provides a mathematically rigorous extension of matrix inversion to singular operators. In the context of PGM, this formulation ensures that the resulting measurement operators remain physically valid and operationally meaningful even in rank-deficient regimes.

\subsection{Inverse vs.\ Pseudoinverse: Conceptual and Operational Difference}

If a Hermitian operator admits the spectral decomposition
\begin{equation}
A = \sum_j \lambda_j \ket{u_j}\bra{u_j},
\end{equation}
then the inverse and pseudoinverse are defined as
\begin{align}
A^{-1} &= \sum_j \lambda_j^{-1} \ket{u_j}\bra{u_j}, \quad \lambda_j \neq 0,\\
A^{+}  &= \sum_{\lambda_j > 0} \lambda_j^{-1} \ket{u_j}\bra{u_j}.
\end{align}

While the ordinary inverse requires all eigenvalues to be nonzero, the pseudoinverse is defined only on the support of $A$, i.e., the subspace corresponding to nonzero eigenvalues. Eigenvectors associated with zero eigenvalues are excluded from the effective dynamics. From an operational perspective, the pseudoinverse corresponds to implementing a \emph{thresholded inverse}, where small eigenvalues are suppressed to avoid numerical instability. This can be expressed through a spectral transformation of the form
\begin{equation}
f(\lambda) =
\begin{cases}
\lambda^{-1}, & \lambda \ge \tau,\\
0, & \lambda < \tau,
\end{cases}
\end{equation}
for some threshold $\tau > 0$.

This interpretation is particularly important in quantum algorithms, where small eigenvalues can lead to large amplification factors and unstable circuit behavior. The threshold therefore acts as a regularization parameter that balances accuracy and stability. Within the framework of block-encoding and spectral-transformation methods, such thresholded spectral mappings can be approximated and subsequently realized through hybrid classical-quantum procedures, providing a practical route toward pseudoinverse-based operations in quantum circuits.

\section{Methodology}\label{methodology}

\subsection{Overall Framework of the Proposed PGM-Based Classification}

\begin{figure*}[t]
\centering
\includegraphics[width=\linewidth]{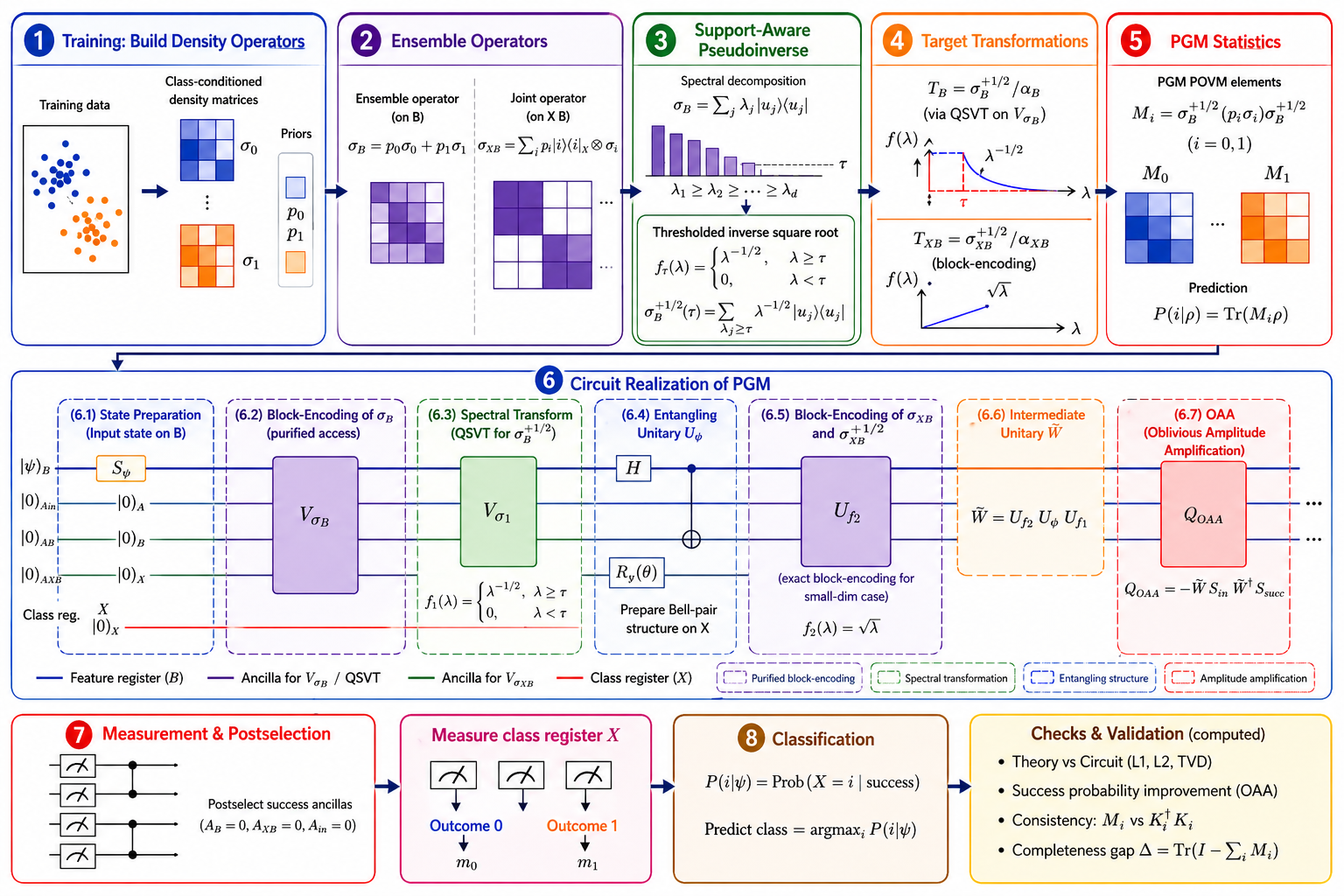}
\caption{Overall framework of the proposed support-aware and threshold-regularized hybrid classical-quantum PGM framework for classification. The pipeline consists of density-operator construction, support-aware pseudoinverse regularization, purified-access block-encoding, spectral transformation, construction of the intermediate unitary $\widetilde{W}$, oblivious amplitude amplification, and final measurement-based classification.}
\label{fig:framework}
\end{figure*}

Fig.~\ref{fig:framework} illustrates the complete workflow of the proposed support-aware and threshold-regularized PGM framework for quantum state discrimination and classification under ill-conditioned ensembles. The procedure begins with a training phase, where input data associated with different classes are encoded into quantum states or density operators. For each class, a representative density operator is constructed, and the prior-weighted ensemble operator is then obtained. This operator captures the global statistical structure of the training ensemble and serves as the central object in the PGM construction. The standard PGM requires the inverse square root of the ensemble operator. However, when the ensemble is rank deficient or ill conditioned, direct inversion becomes unstable or undefined. To address this, we introduce a support-aware and threshold-regularized pseudoinverse construction. The support-aware step restricts the measurement to the effective subspace of the ensemble operator, while threshold regularization suppresses eigenvalues below a prescribed cutoff. This yields a stable pseudoinverse square-root transformation and ensures that the resulting measurement remains physically meaningful even in singular or nearly singular regimes. Beyond the matrix-level formulation, the proposed framework provides a hybrid classical-quantum circuit realization. In the final implementation, the PGM protocol is realized through purified-access block-encodings of the relevant operators, spectral transformation of the ensemble operator, construction of an intermediate unitary $\widetilde{W}$, and an oblivious amplitude amplification (OAA) step. The measurement of the class register produces the predicted label. This design bridges the gap between the theoretical PGM definition and an executable quantum circuit.

\subsection{PGM Specialization for Class-Conditioned Quantum Ensembles}

We consider a supervised classification setting with class label register $X$ and feature-state register $B$. For a binary problem, $X$ is a single qubit and $B$ is a single feature qubit. Let $\sigma_0$ and $\sigma_1$ denote the class-conditioned density operators on register $B$, and let $p_0$ and $p_1$ denote the corresponding class priors. The joint classical-quantum training state is written as
\begin{equation}
\sigma_{XB}
=
p_0 \ket{0}\bra{0}_X \otimes \sigma_0
+
p_1 \ket{1}\bra{1}_X \otimes \sigma_1 .
\end{equation}
The marginal ensemble operator on the feature register is obtained by tracing out the class register,
\begin{equation}
\sigma_B = \mathrm{Tr}_X(\sigma_{XB})
= p_0\sigma_0 + p_1\sigma_1 .
\end{equation}
The PGM measurement operators are then
\begin{equation}
M_i
=
\sigma_B^{+1/2}\,
(p_i\sigma_i)\,
\sigma_B^{+1/2},
\end{equation}
where $\sigma_B^{+1/2}$ denotes the square root of the Moore-Penrose pseudoinverse of $\sigma_B$. Equivalently, the full PGM structure can be expressed through the class-feature operator $\sigma_{XB}$ and the marginal operator $\sigma_B$. This form is particularly convenient for circuit implementation because it separates the procedure into two transformations: one acting on $\sigma_B$ and another acting on $\sigma_{XB}$.

\subsection{Support-Aware Threshold-Regularized Pseudoinverse}

For the PGM case, the desired spectral map is
\begin{equation}
f_{\tau}(\lambda) =
\begin{cases}
\lambda^{-1/2}, & \lambda \ge \tau,\\
0, & \lambda < \tau,
\end{cases}
\end{equation}
where $\tau>0$ is a conditioning threshold. Given the spectral decomposition
\begin{equation}
\sigma_B = \sum_j \lambda_j \ket{u_j}\bra{u_j},
\end{equation}
the threshold-regularized pseudoinverse square root is defined as
\begin{equation}
\sigma_B^{+1/2}(\tau)
=
\sum_{\lambda_j\ge \tau}
\lambda_j^{-1/2}
\ket{u_j}\bra{u_j}.
\end{equation}
This construction removes eigencomponents associated with eigenvalues below the threshold and therefore avoids the numerical instability caused by small eigenvalues. In the full-rank and well-conditioned case, this reduces to the ordinary inverse square root. In rank-deficient cases, it acts only on the support of $\sigma_B$ and annihilates the null space. The threshold $\tau$ therefore plays the role of a regularization parameter. Small values of $\tau$ retain more spectral information but may amplify noise, while larger values improve conditioning at the cost of discarding weak spectral components. This trade-off is analyzed experimentally in the results section.

\subsection{Quantum Circuit Realization via Block-Encoding and Spectral Transformation}

The HHL algorithm~\cite{harrow2009quantum} provides a foundational framework for implementing spectral transformations of matrices in quantum computing. By combining quantum phase estimation with controlled rotations, HHL enables eigenvalue-dependent transformations such as $\lambda^{-1}$ to be encoded into quantum amplitudes. Recent works have explored extensions of such spectral techniques to improve robustness in the presence of ill-conditioned operators, including strategies for suppressing contributions from small eigenvalues~\cite{CarreraVazquez2021QLSA}. In the present work, we do not rely on a direct HHL implementation. Instead, we adopt a block-encoding and spectral-transformation viewpoint, which is better suited to a constructive circuit realization of PGM. The transformed operator $\sigma_B^{+1/2}$ is represented through a QSVT-style spectral-transformation procedure applied to a purified-access block-encoding of $\sigma_B$. The required phase parameters are obtained through classical optimization, and the resulting transformation is subsequently realized and validated at the circuit level. This provides a circuit-level representation and validation of the thresholded pseudoinverse square-root operation required by PGM. Similarly, the operator $\sigma_{XB}^{1/2}$ is realized as a block-encoded transformation on the joint class-feature register. In the final implementation, the block-encoding of $\sigma_{XB}^{1/2}$ is constructed exactly for the small-dimensional binary case, while the inverse-square-root transformation of $\sigma_B$ is obtained through optimized phase factors in the QSVT-style sequence. This preserves the constructive structure of the reference approach while allowing explicit numerical validation at the circuit level.

\subsection{Purified-Access Block-Encoding of $\sigma_B$ and $\sigma_{XB}$}

The first circuit-level step is to construct purified states corresponding to $\sigma_B$ and $\sigma_{XB}$. For a density operator $\rho$ with spectral decomposition
\begin{equation}
\rho = \sum_j \lambda_j \ket{u_j}\bra{u_j},
\end{equation}
a purification can be written as
\begin{equation}
\ket{\psi_\rho}
=
\sum_j \sqrt{\lambda_j}\ket{j}_{R}\ket{u_j}_{S},
\end{equation}
where $R$ is a reference register and $S$ is the system register. A state-preparation unitary $U_\rho$ is constructed such that
\begin{equation}
U_\rho\ket{0}_{RS} = \ket{\psi_\rho}.
\end{equation}
Using $U_\rho$, an exact purified-access block-encoding of $\rho$ is obtained by applying $U_\rho$, swapping the system register with an external system register, and then applying $U_\rho^\dagger$. Operationally, this realizes a unitary $V_\rho$ whose appropriate zero-ancilla block equals $\rho$:
\begin{equation}
\bra{0}_{A} V_\rho \ket{0}_{A} = \rho .
\end{equation}
In the implementation, this procedure is applied separately to $\sigma_B$ and $\sigma_{XB}$, yielding block-encodings $V_{\sigma_B}$ and $V_{\sigma_{XB}}$. The final code verifies these constructions by explicitly extracting the corresponding zero-ancilla blocks and comparing them with the target matrices.

\subsection{Construction of the Transformed Blocks}

After obtaining the block-encodings, the next step is to construct the transformed operators required by the PGM. For $\sigma_B$, the target transformation is
\begin{equation}
T_B = \frac{\sigma_B^{+1/2}}{\alpha_B},
\end{equation}
where $\alpha_B$ is a normalization factor chosen so that $\|T_B\|\le 1$. The circuit implements this transformation through a QSVT-style sequence of alternating block-encoding and phase-reflection operations:
\begin{equation}
U_{f_1}
=
R_{\phi_m}
V_{\sigma_B}^{(\dagger)}
R_{\phi_{m-1}}
\cdots
V_{\sigma_B}
R_{\phi_0},
\end{equation}
where the phases $\{\phi_j\}$ are optimized so that the extracted block of $U_{f_1}$ approximates $T_B$. For the joint class-feature operator, the target transformation is
\begin{equation}
T_{XB} = \frac{\sigma_{XB}^{1/2}}{\alpha_{XB}},
\end{equation}
where $\alpha_{XB}$ ensures contraction. In the small-dimensional setting considered here, this transformation is implemented as an exact block-encoding. Hence, the extracted zero-ancilla blocks satisfy
\begin{equation}
\bra{0}_{A_B} U_{f_1} \ket{0}_{A_B}
\approx
\frac{\sigma_B^{+1/2}}{\alpha_B},
\end{equation}
and
\begin{equation}
\bra{0}_{A_{XB}} U_{f_2} \ket{0}_{A_{XB}}
=
\frac{\sigma_{XB}^{1/2}}{\alpha_{XB}}.
\end{equation}

\subsection{Construction of the Intermediate Unitary $\widetilde{W}$}

The transformed blocks are combined to form the intermediate unitary $\widetilde{W}$, which realizes the PGM measurement statistics after postselection on the success ancillas. In the binary implementation, the global circuit uses a class register $X$, a feature register $B$, and ancilla registers associated with the block-encodings. A Bell-pair preparation unitary $U_\phi$ is used to introduce the required entangled structure between the class register and an auxiliary register. The circuit-level construction has the form
\begin{equation}
\widetilde{W}
=
U_{f_2}\,
U_{\phi}\,
U_{f_1},
\end{equation}
where $U_{f_1}$ implements the transformed block associated with $\sigma_B^{+1/2}$, $U_\phi$ prepares the required entangled class-register structure, and $U_{f_2}$ implements the transformed block associated with $\sigma_{XB}^{1/2}$. In the final implementation, careful register ordering is essential. The correct class-feature ordering is determined by extracting the zero-ancilla block and verifying that it reproduces $\sigma_{XB}$ with numerical precision. This register-alignment step is crucial for obtaining the correct PGM probabilities. After applying $\widetilde{W}$ to an input feature state and postselecting on the success ancillas, measuring the class register yields probabilities
\begin{equation}
\Pr(i|\psi)
=
\mathrm{Tr}(M_i \ket{\psi}\bra{\psi}),
\end{equation}
which match the theoretical PGM statistics.

\subsection{Oblivious Amplitude Amplification}

The success probability of the postselected branch can be small because the desired transformed block appears only in the zero-ancilla subspace. To improve practical feasibility, we apply an oblivious amplitude amplification step. Let $P_{\mathrm{in}}$ denote the projector onto the valid input subspace and let $P_{\mathrm{succ}}$ denote the projector onto the success-ancilla subspace. The corresponding reflections are
\begin{equation}
S_{\mathrm{in}} = I - 2P_{\mathrm{in}},
\end{equation}
and
\begin{equation}
S_{\mathrm{succ}} = I - 2P_{\mathrm{succ}}.
\end{equation}
The one-step OAA operator is constructed as
\begin{equation}
Q_{\mathrm{OAA}}
=
-\widetilde{W}
S_{\mathrm{in}}
\widetilde{W}^{\dagger}
S_{\mathrm{succ}}.
\end{equation}
Importantly, the amplification operator is applied to the state $\widetilde{W}\ket{\psi}$ rather than directly to the original input state. Thus, the amplified state is computed as
\begin{equation}
\ket{\Psi_{\mathrm{amp}}}
=
Q_{\mathrm{OAA}}\widetilde{W}\ket{\psi}.
\end{equation}
This preserves the PGM output probabilities while increasing the probability of observing the success branch. In the final circuit-level implementation, OAA significantly increases the success probability while preserving the output class probabilities.

\subsection{From a Pseudoinverse Circuit to the PGM POVM Elements}
The following discussion provides an operational interpretation of the proposed framework and is not intended as a claim that the pseudoinverse primitive is obtained through a fully quantum procedure. Assume that a quantum circuit primitive is available that coherently implements the pseudoinverse square-root of the ensemble operator, namely a block that realizes approximately, and on the support of $S$,
\begin{equation}
U_{S^{+1/2}}:\quad \ket{\psi}\ket{0}_{a}\;\longmapsto\;
\Big( S^{+1/2}\ket{\psi}\Big)\ket{1}_{a}\;+\;\ket{\Phi^\perp}\ket{0}_{a},
\end{equation}
where $a$ is an ancilla flag qubit and $\ket{\Phi^\perp}$ collects failure components. The goal is to implement the PGM POVM elements
\begin{equation}
M_i \;=\; S^{+1/2}\,(p_i\rho_i)\,S^{+1/2}.
\end{equation}
A POVM $\{M_i\}$ is defined through outcome statistics,
\begin{equation}
\Pr(i \mid \sigma) = \mathrm{Tr}(M_i \sigma).
\end{equation}
Thus, implementing the POVM does not require classically materializing the matrices $M_i$. It is sufficient to construct a quantum instrument whose measurement statistics reproduce these probabilities. Any POVM admits a Naimark dilation,
\begin{equation}
M_i = V^\dagger (I \otimes \ket{i}\bra{i}) V,
\end{equation}
for some isometry $V$. For PGM, define
\begin{equation}
K_i := \sqrt{p_i}\,\sqrt{\rho_i}\,S^{+1/2}.
\end{equation}
Then
\begin{equation}
K_i^\dagger K_i = M_i.
\end{equation}
The isometry
\begin{equation}
V:\quad \ket{\psi} \mapsto \sum_i \ket{i}\otimes K_i \ket{\psi}
\end{equation}
therefore implements the PGM measurement. This provides an operational interpretation of the proposed circuit construction: the circuit realizes the PGM through an isometric embedding, followed by measurement of the class register.

\subsection{Numerical PGM Construction}

\begin{algorithm}[!t]
\DontPrintSemicolon
\nonl \textbf{Input:} Priors $\{p_i\}$, states $\{\rho_i\}$, method $\in$ \{\texttt{inverse}, \texttt{pseudoinverse}\}, threshold $\tau$\\
\nonl \textbf{Output:} PGM elements $\{M_i\}$, success probability $P_{\mathrm{succ}}$, trace gap $\Delta$

\textbf{Compute:} Ensemble operator
\[
S = \sum_i p_i \rho_i
\]

\textbf{Eigen-decompose:} $S = \sum_j \lambda_j \ket{u_j}\bra{u_j}$\;

\eIf{method = inverse \textbf{and} $\lambda_j \ge \tau$}{
    $T \leftarrow \sum_j \lambda_j^{-1/2} \ket{u_j}\bra{u_j}$\;
}{
    $T \leftarrow \sum_{\lambda_j \ge \tau} \lambda_j^{-1/2} \ket{u_j}\bra{u_j}$\;
}

\ForEach{$i=1,\dots,K$}{
    $M_i \leftarrow T (p_i \rho_i) T$\;
    $M_i \leftarrow \frac{1}{2}(M_i + M_i^\dagger)$\;
}

\textbf{Compute:}
\[
\Delta = \mathrm{Tr}\!\left(I - \sum_i M_i\right), \quad
P_{\mathrm{succ}} = \sum_i p_i\, \mathrm{Tr}(M_i \rho_i)
\]

\textbf{Return:} $\{M_i\}$, $P_{\mathrm{succ}}$, $\Delta$

\caption{Pseudoinverse-Based Pretty Good Measurement}
\label{alg:pgm_pseudoinverse}
\end{algorithm}

Algorithm~\ref{alg:pgm_pseudoinverse} summarizes the numerical construction of the threshold-regularized pseudoinverse PGM. The procedure begins with an input ensemble of prior probabilities $\{p_i\}_{i=1}^{K}$ and corresponding quantum states $\{\rho_i\}_{i=1}^{K}$. From this ensemble, the average state $S$ is constructed. The spectral decomposition of $S$ is then used to compute either the ordinary inverse square root or the thresholded pseudoinverse square root. If the inverse branch is used, all eigenvalues must be sufficiently large. If this condition fails, the inverse-based construction becomes unstable. In contrast, the pseudoinverse branch discards eigenvalues below $\tau$ and constructs the inverse square root only on the effective support of $S$. Once the transformation operator $T$ is obtained, the PGM elements are built as
\begin{equation}
M_i = T\,(p_i \rho_i)\,T.
\end{equation}
A Hermitian symmetrization step removes small numerical asymmetries. The completeness of the resulting POVM is evaluated through the trace gap
\begin{equation}
\Delta = \mathrm{Tr}\!\left(I-\sum_i M_i\right).
\end{equation}
For full-rank ensembles, this gap is approximately zero, whereas for thresholded or rank-deficient cases it reflects the portion of the Hilbert space removed from the effective support. The success probability is computed as
\begin{equation}
P_{\mathrm{succ}}=\sum_{i=1}^{K} p_i\, \mathrm{Tr}(M_i \rho_i),
\end{equation}
which provides the main quantitative measure of discrimination performance. In addition, the Kraus-operator representation
\begin{equation}
K_i = \sqrt{p_i}\,\sqrt{\rho_i}\,S^{+1/2}
\end{equation}
is used as a consistency check by comparing $\widetilde{M}_i=K_i^\dagger K_i$ with the directly computed $M_i$.

\subsection{Experimental Evaluation Pipeline}

\begin{algorithm}[!t]
\DontPrintSemicolon
\nonl \textbf{Input:} Thresholds $\{\tau\}$, noise levels $\{\lambda\}$, class counts $\{K\}$, ensemble generators, circuit construction routine\\
\nonl \textbf{Output:} Tables and plots for threshold, noise, scaling, circuit validation, and consistency

\textbf{Initialize:} Define ensembles and parameter grids\;

\BlankLine
\textbf{Threshold Sweep:}
Generate ill-conditioned ensemble\;
\ForEach{$\tau$}{
    Apply Algorithm~\ref{alg:pgm_pseudoinverse}\;
    Record rank, $P_{\mathrm{succ}}$, $\Delta$\;
}

\BlankLine
\textbf{Noise Robustness:}
\ForEach{$\lambda$}{
    Apply depolarization and compute $P_{\mathrm{succ}}$\;
}

\BlankLine
\textbf{Class Scaling:}
\ForEach{$K$}{
    Generate ensemble and compute $P_{\mathrm{succ}}$\;
}

\BlankLine
\textbf{Circuit-Level Validation:}
Construct $V_{\sigma_B}$, $V_{\sigma_{XB}}$, $U_{f_1}$, $U_{f_2}$, $\widetilde{W}$, and OAA\;
Compare theoretical and circuit-generated probabilities using L1, L2, and TVD distances\;

\BlankLine
\textbf{Consistency Check:}
Compute $\delta_i = \|M_i - \widetilde{M}_i\|_F$ for all $i$\;

\BlankLine
\textbf{Structural Summary:}
Evaluate full-rank, rank-deficient, and ill-conditioned cases\;

\textbf{Return:} All recorded results and summaries\;

\caption{Experimental Evaluation Pipeline for Robust PGM}
\label{alg:pgm_experimental_pipeline}
\end{algorithm}

Algorithm~\ref{alg:pgm_experimental_pipeline} summarizes the complete experimental workflow used to evaluate the proposed framework. While Algorithm~\ref{alg:pgm_pseudoinverse} describes the construction of the measurement operators for a fixed ensemble, Algorithm~\ref{alg:pgm_experimental_pipeline} organizes repeated evaluations across threshold values, noise levels, class counts, and circuit-level implementations. The first block performs a threshold sweep on an ill-conditioned ensemble. This quantifies how the retained effective rank, success probability, and trace gap vary as the threshold $\tau$ changes. The second block studies robustness under depolarizing noise by mixing each state with the maximally mixed state. The third block investigates scaling with the number of classes. The fourth block performs the circuit-level validation central to this work: the purified block-encodings, transformed blocks, intermediate unitary $\widetilde{W}$, and OAA procedure are constructed explicitly and compared against theoretical PGM probabilities. The agreement is measured using L1 distance, L2 distance, and total variation distance (TVD). Finally, the Kraus consistency check verifies that the direct matrix-level PGM construction agrees with the operational quantum-instrument representation. Together, these procedures provide a comprehensive validation of the proposed robust PGM framework at both the matrix and circuit levels. The proposed approach does not constitute a fully quantum computation of the Moore–Penrose pseudoinverse. Spectral information, polynomial synthesis, and transformation parameters are obtained through classical preprocessing, while the resulting operators are realized and validated within a quantum-circuit framework. Consequently, the present method should be interpreted as a hybrid classical–quantum realization of pseudoinverse-based PGM rather than a fully quantum implementation.

\section{Experimental Results}\label{results}

\subsection{Experimental Setup}

All experiments were carried out using Python in a Google Colab environment. The study combines matrix-level validation, synthetic robustness analysis, and end-to-end circuit-level simulation. Matrix-level experiments were used to evaluate the behavior of the support-aware pseudoinverse PGM under well-conditioned, rank-deficient, ill-conditioned, noisy, and multi-class regimes. For these experiments, standard numerical linear algebra routines were used for spectral decomposition, inverse square-root evaluation, and Moore-Penrose pseudoinverse construction. For the circuit-level experiments, we implemented a hybrid classical-quantum PGM pipeline using purified-access block-encodings, QSVT-style spectral transformations, the intermediate unitary $\widetilde{W}$, and oblivious amplitude amplification. The final circuit was evaluated on a real binary Lupus dataset. The main validation criterion was the agreement between the theoretical PGM probabilities and the circuit-generated probabilities, quantified using L1 distance, L2 distance, and TVD. Classification accuracy was also reported, but the primary objective of the circuit experiment was to verify faithful measurement realization rather than maximize classical predictive performance.

\subsection{Datasets}

Two complementary categories of data were considered in this work to validate the proposed support-aware pseudoinverse-based PGM framework at both the operator and circuit levels. For matrix-level analysis, we constructed synthetic quantum ensembles designed to probe different spectral regimes of the ensemble operator $S$. In particular, we considered well-conditioned (full-rank), rank-deficient, and ill-conditioned ensembles. These controlled instances enable systematic investigation of the behavior of inverse and pseudoinverse constructions, the role of threshold regularization, and the consistency between direct POVM and Kraus-operator realizations. Since these experiments focus on the intrinsic properties of the PGM construction, synthetic ensembles provide a clean and interpretable setting in which spectral characteristics can be precisely controlled. 

To evaluate the proposed framework on realistic data, we additionally considered a real-world Lupus image dataset obtained from Kaggle~\cite{lupus_dataset}. The dataset consists of dermoscopic skin images labeled as Lupus and non-Lupus, comprising a total of $590$ samples, including $311$ Lupus images and $279$ non-Lupus images. The dataset was divided into $442$ training samples and $148$ test samples using stratified sampling to preserve the class distribution across both subsets. Due to current quantum resource limitations and the objective of obtaining an explicit circuit-level realization, each image was represented by a two-dimensional feature vector consisting of the grayscale intensity mean and standard deviation. The extracted features were normalized using min-max scaling and subsequently amplitude encoded into single-qubit quantum states. These encoded quantum states were used to construct class density operators, estimate the ensemble operator, and generate the corresponding threshold-regularized PGM elements. The primary purpose of the Lupus experiment is not to compete with state-of-the-art classical image-classification pipelines, but rather to validate the proposed pseudoinverse-based PGM framework on a realistic dataset. In particular, the experiment is designed to assess the agreement between theoretical PGM probabilities and those obtained through explicit circuit-level realizations based on block-encoding and QSVT-style spectral transformations. This setting provides a practical demonstration of the proposed hybrid classical-quantum framework and allows evaluation of its numerical stability, robustness, and fidelity under realistic data conditions. Throughout the experimental study, synthetic ensembles were used to investigate the mathematical properties of the proposed support-aware pseudoinverse formulation, while the Lupus dataset served as a realistic benchmark for validating the circuit-level realization and classification behavior of the framework.

\subsection{Experimental Evaluation}
\label{sec:experiments}

In this section, we numerically validate the proposed pseudoinverse-based PGM framework across several representative regimes: a well-conditioned binary ensemble, a rank-deficient ensemble, an ill-conditioned ensemble, a threshold sweep for the pseudoinverse square-root, robustness under depolarizing noise, scaling with the number of classes, and consistency between the direct PGM construction and its Kraus-operator realization. The central objective is to demonstrate that while the ordinary inverse-based construction is only valid for well-conditioned full-rank ensemble operators, the pseudoinverse-based formulation remains operationally meaningful and numerically stable for singular and nearly singular cases. The experiments were implemented in Google Colab using numerical spectral decomposition and matrix-based simulation of the PGM operators.

\subsubsection{Well-conditioned binary ensemble}
We first consider a well-conditioned binary qubit ensemble for which the ensemble average state $S$ is full rank. In this regime, both the ordinary inverse and the pseudoinverse are expected to coincide. The numerical results are reported in Table~\ref{tab:exp1_well_conditioned}. The minimum and maximum eigenvalues of the ensemble operator are $0.038060$ and $0.961940$, respectively, indicating that $S$ is positive definite and invertible. As expected, both the inverse-based and pseudoinverse-based PGM constructions succeed and produce identical average success probabilities of $0.691342$. Moreover, the trace gap $\mathrm{Tr}\!\left(I - \sum_i M_i\right)$ is numerically zero for both constructions, confirming exact completeness up to floating-point precision. The corresponding ensemble average state is
\begin{equation}
S =
\begin{bmatrix}
0.9268 & 0.1768 \\
0.1768 & 0.0732
\end{bmatrix},
\end{equation}
which is clearly full rank. This experiment serves as a sanity check for the implementation and confirms that the pseudoinverse-based construction reduces to the standard PGM whenever the ensemble operator is well-conditioned.

\begin{table*}[htbp]
\centering
\caption{Results for the well-conditioned binary qubit ensemble.}
\label{tab:exp1_well_conditioned}
\begin{tabular}{lcccccc}
\hline\hline
\textbf{Case} & \textbf{dim} & \textbf{$K$} & \textbf{rank$(S)$} & \textbf{min eig$(S)$} & \textbf{max eig$(S)$} & \textbf{Inverse Status} \\
\hline
Well-conditioned binary qubit & 2 & 2 & 2 & 0.038060 & 0.961940 & OK \\
\hline
\textbf{Case} & \textbf{Inverse Success} & \textbf{Inverse TraceGap} & \textbf{Pseudo Status} & \textbf{Pseudo Success} & \textbf{Pseudo TraceGap} & \\
\hline
Well-conditioned binary qubit & 0.691342 & 0.000000 & OK & 0.691342 & 0.000000 & \\
\hline
\end{tabular}
\end{table*}

\subsubsection{Rank-deficient ensemble}
We next study a $4$-dimensional ensemble consisting of three states whose support lies entirely within a $2$-dimensional subspace. In this case, the ensemble operator is rank deficient, with eigenvalues
\begin{equation}
\mathrm{eig}(S) = [0,\;0,\;0.3,\;0.7],
\end{equation}
and hence $\mathrm{rank}(S)=2$. The explicit matrix form is
\begin{equation}
S =
\begin{bmatrix}
0.5 & 0.2 & 0   & 0 \\
0.2 & 0.5 & 0   & 0 \\
0   & 0   & 0   & 0 \\
0   & 0   & 0   & 0
\end{bmatrix}.
\end{equation}

As shown in Table~\ref{tab:exp2_rank_deficient}, the inverse-based construction fails, since the inverse square root of $S$ is undefined due to zero eigenvalues. In contrast, the pseudoinverse-based construction remains well defined and yields a success probability of $0.639253$. The reported trace gap $\mathrm{Tr}\!\left(I - \sum_i M_i\right) = 2.0$ is not a defect of the method; rather, it reflects the fact that the pseudoinverse-based PGM is complete on the support of $S$ rather than on the full $4$-dimensional Hilbert space. Since the null space of $S$ has dimension $2$, the trace gap equals the dimension of the discarded subspace. This result directly confirms the theoretical claim that the Moore-Penrose pseudoinverse is the correct generalization of the inverse in rank-deficient PGM settings.

\begin{table}[t]
\centering
\caption{Results for the rank-deficient ensemble.}
\label{tab:exp2_rank_deficient}
\begin{tabular}{lcccc}
\hline\hline
\textbf{Case} & \textbf{dim} & \textbf{$K$} & \textbf{rank$(S)$} & \textbf{eig range} \\
\hline
Rank-deficient 4D & 4 & 3 & 2 & $[0,\;0.7]$ \\
\hline
\end{tabular}

\vspace{0.3em}

\begin{tabular}{lccc}
\hline
\textbf{Method} & \textbf{Status} & \textbf{Success} & \textbf{Trace Gap} \\
\hline
Inverse      & Failed & -       & - \\
Pseudoinverse & OK     & 0.639253 & 2.000000 \\
\hline
\end{tabular}
\end{table}
\subsubsection{Ill-conditioned ensemble}
We further consider a nearly singular $4$-dimensional ensemble, designed so that the ensemble operator is formally full rank but extremely ill conditioned. The smallest eigenvalue is approximately $1.25\times 10^{-11}$, while the largest eigenvalue is $0.796535$, which makes direct inversion numerically unstable. The eigenvalue spectrum is
\begin{equation}
\mathrm{eig}(S) = [0,\;0.0785,\;0.125,\;0.7965],
\end{equation}
where the smallest eigenvalue is numerically close to zero at the chosen tolerance level. The results in Table~\ref{tab:exp3_ill_conditioned} show that the inverse-based method again fails due to near-zero eigenvalues, whereas the pseudoinverse-based construction succeeds and yields an average success probability of $0.597535$. The trace gap is $1.0$, indicating that one effective spectral component is removed by thresholding, so the implemented PGM remains complete on a $3$-dimensional support subspace. This experiment highlights the practical relevance of thresholded pseudoinverse constructions: even when the operator is technically invertible, very small eigenvalues make the ordinary inverse unstable, whereas the pseudoinverse yields a robust and physically meaningful measurement.

\begin{table}[t]
\centering
\caption{Results for the ill-conditioned ensemble.}
\label{tab:exp3_ill_conditioned}
\begin{tabular}{lcccc}
\hline\hline
\textbf{Case} & \textbf{dim} & \textbf{$K$} & \textbf{rank$(S)$} & \textbf{eig range} \\
\hline
Ill-conditioned & 4 & 4 & 3 & $[0,\;0.7965]$ \\
\hline
\end{tabular}

\vspace{0.3em}

\begin{tabular}{lccc}
\hline
\textbf{Method} & \textbf{Status} & \textbf{Success} & \textbf{Trace Gap} \\
\hline
Inverse       & Failed & -       & - \\
Pseudoinverse & OK     & 0.597535 & 1.000000 \\
\hline
\end{tabular}
\end{table}
\subsubsection{Threshold sweep for the pseudoinverse square-root}
\begin{figure*}[t]
\centering

\begin{subfigure}{0.49\linewidth}
\centering
\includegraphics[width=\linewidth]{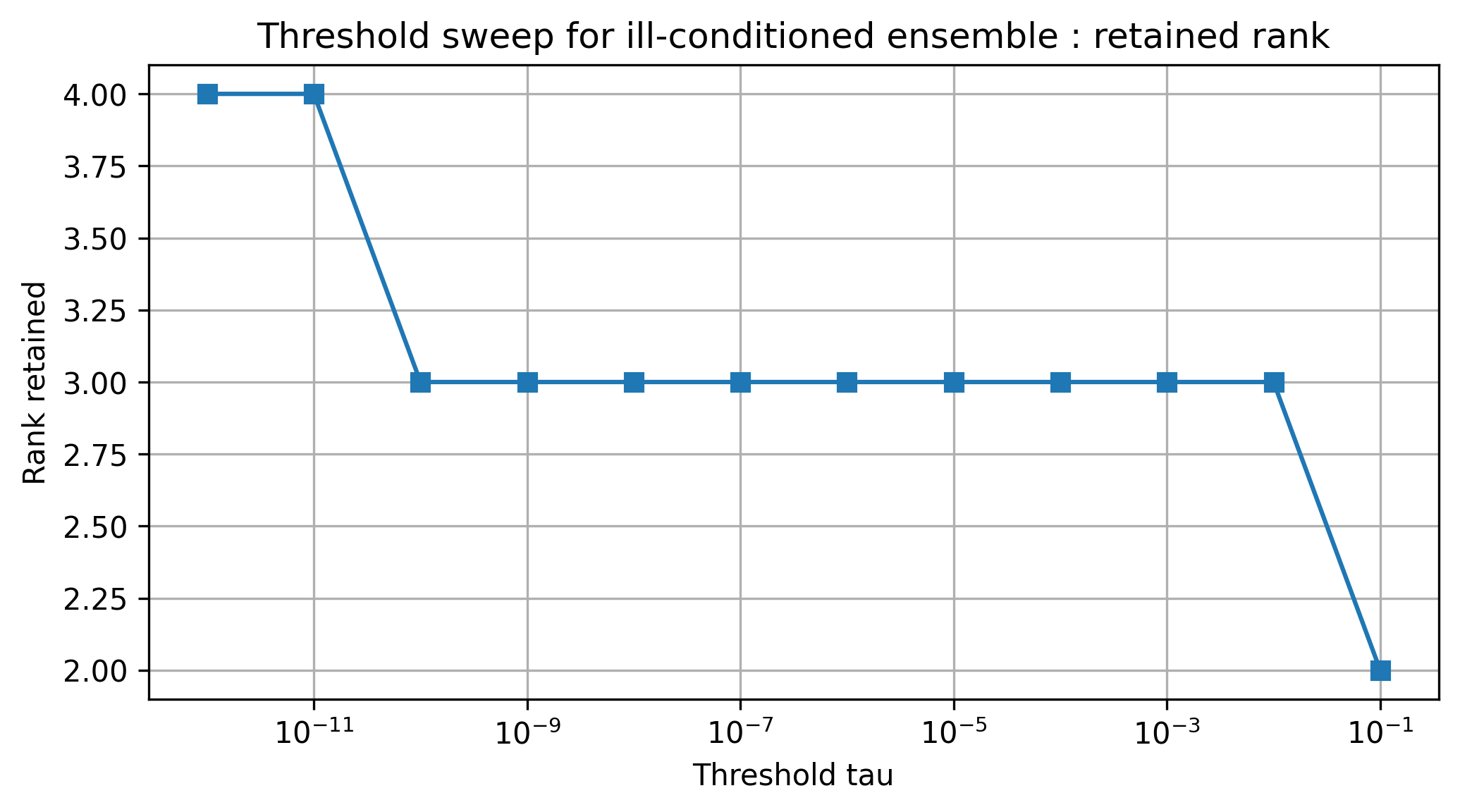}
\caption{Retained effective rank vs threshold $\tau$.}
\label{fig:threshold_rank}
\end{subfigure}
\hfill
\begin{subfigure}{0.49\linewidth}
\centering
\includegraphics[width=\linewidth]{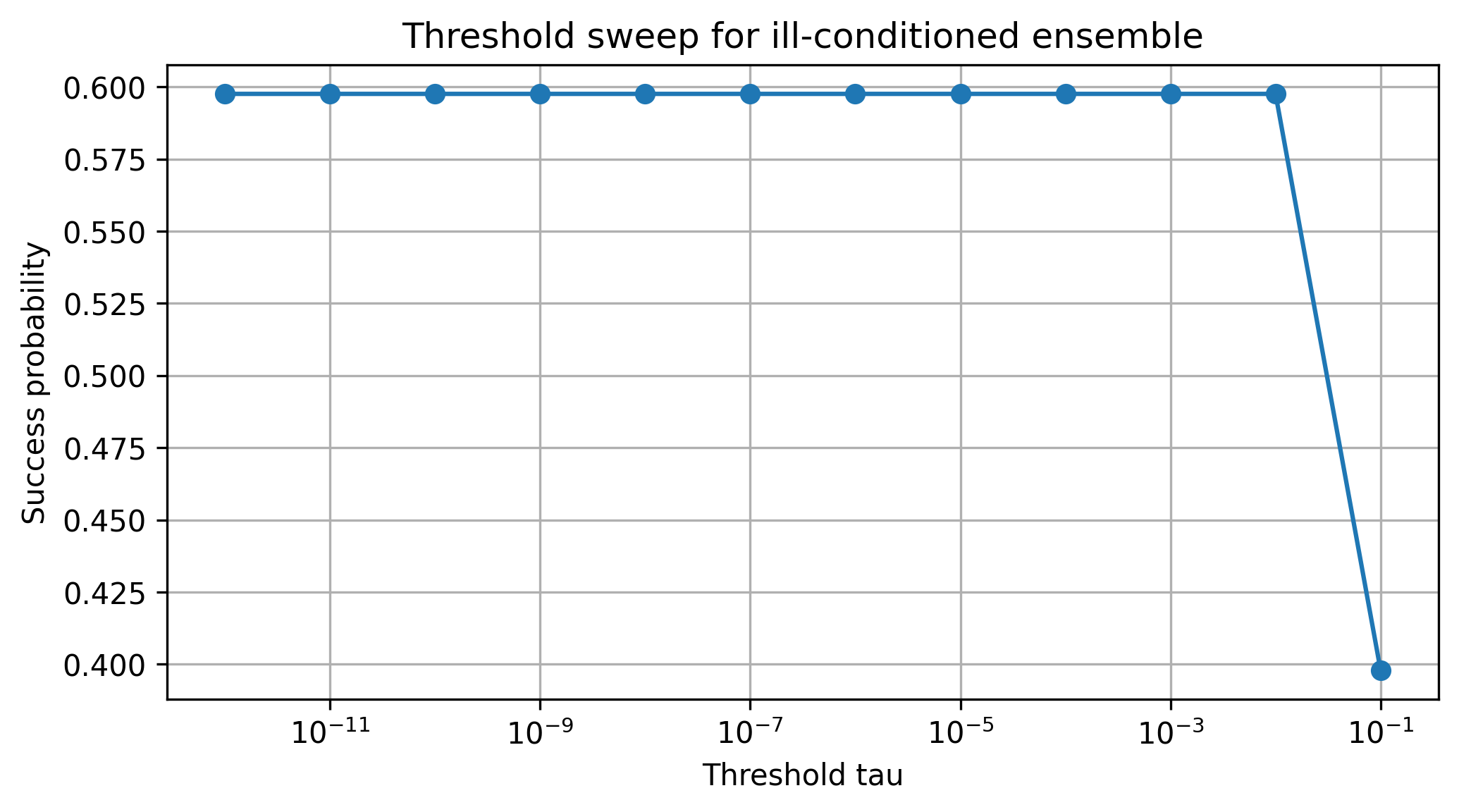}
\caption{Success probability vs threshold $\tau$.}
\label{fig:threshold_success}
\end{subfigure}

\vspace{0.5em}

\begin{subfigure}{0.49\linewidth}
\centering
\includegraphics[width=\linewidth]{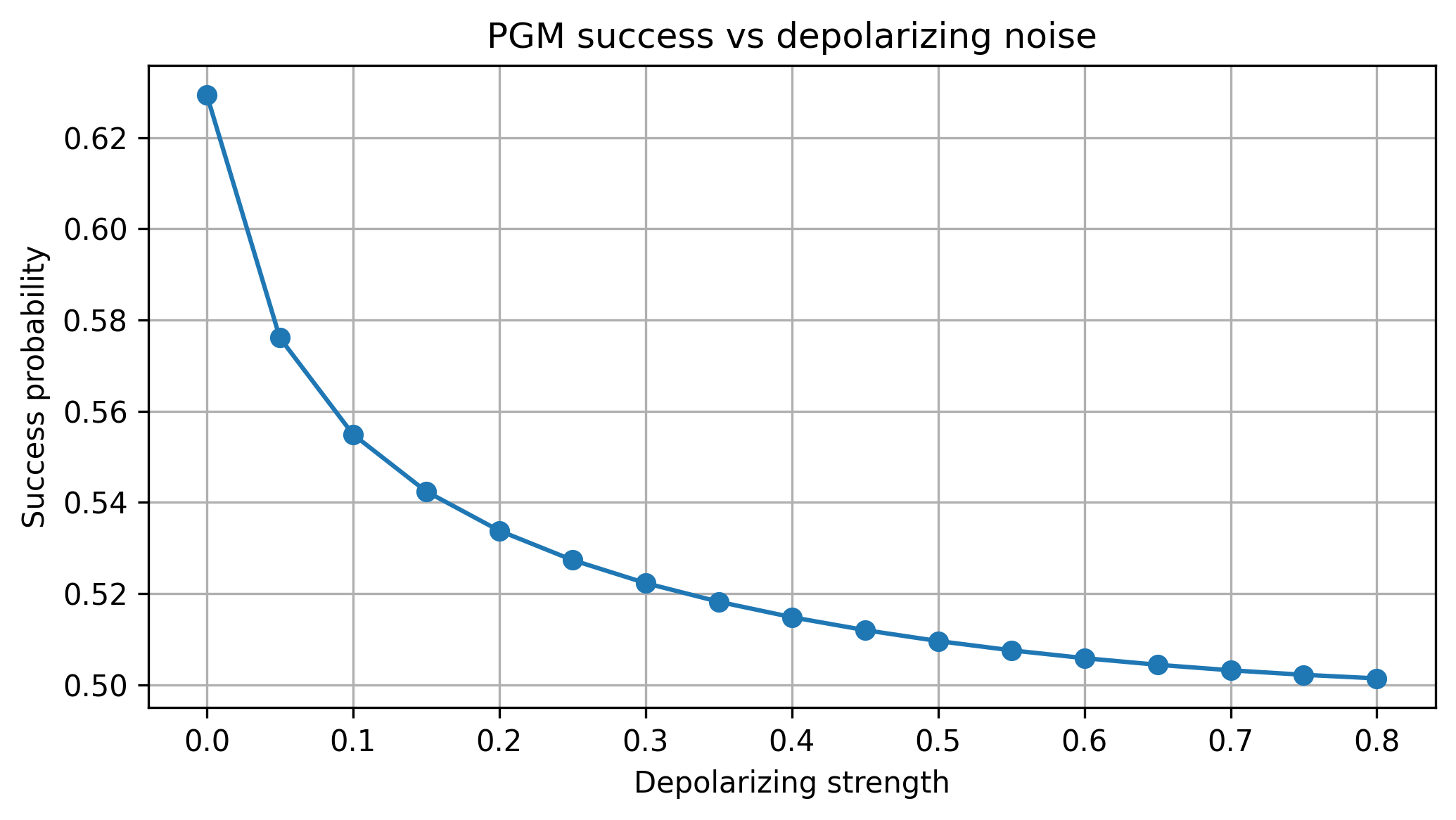}
\caption{Success probability under depolarizing noise.}
\label{fig:noise_robustness}
\end{subfigure}
\hfill
\begin{subfigure}{0.49\linewidth}
\centering
\includegraphics[width=\linewidth]{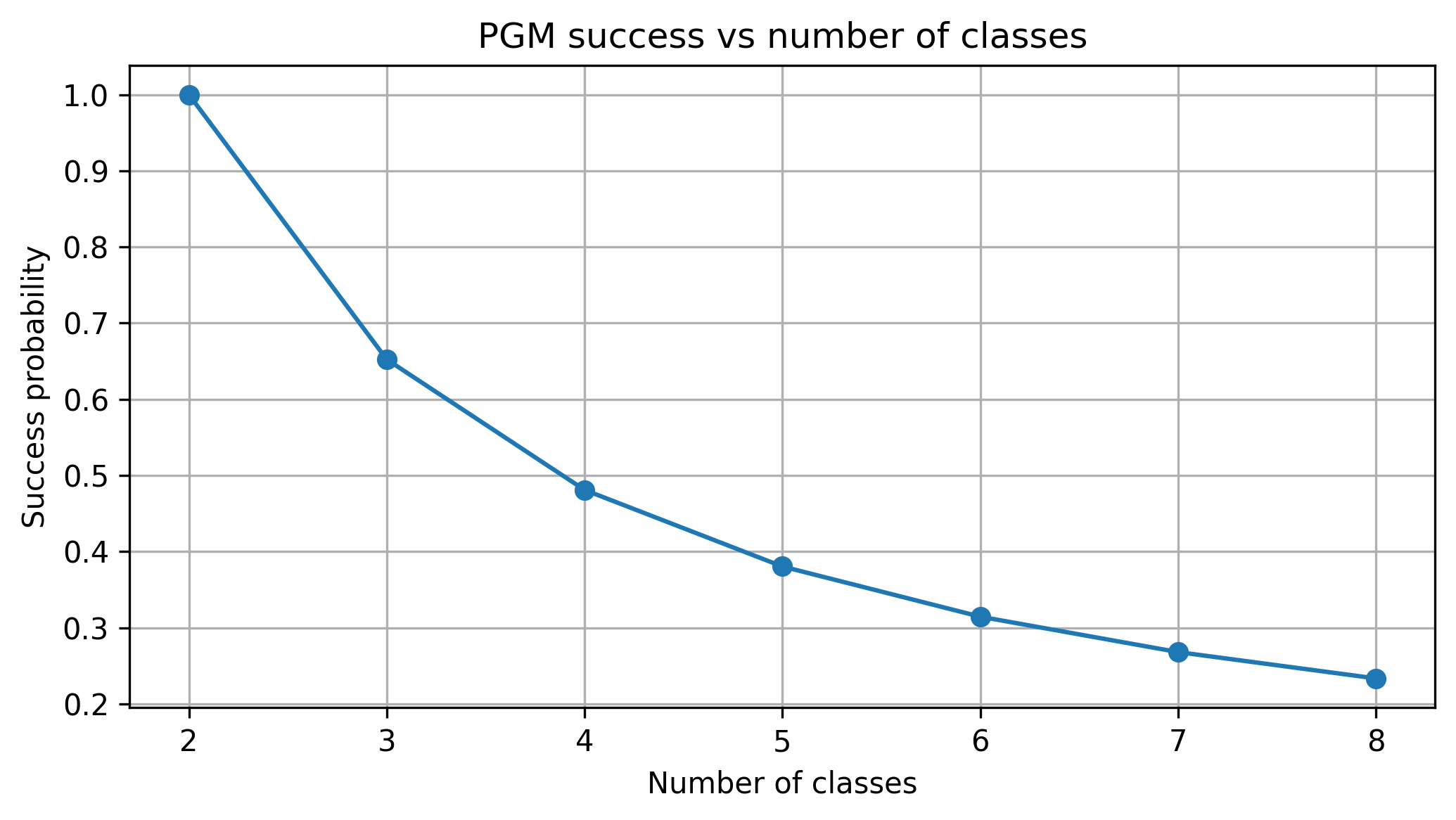}
\caption{Success probability vs number of classes.}
\label{fig:class_scaling}
\end{subfigure}

\caption{
Performance analysis of the proposed pseudoinverse-based PGM under different conditions.
(a) Retained effective rank as a function of threshold $\tau$, showing removal of unstable eigenvalues.
(b) Corresponding success probability, demonstrating robustness across a wide threshold range.
(c) Robustness under depolarizing noise, where performance degrades with increasing noise strength.
(d) Scaling behavior with respect to the number of classes, indicating increased difficulty in state discrimination for larger ensembles.
}
\label{fig:pgm_performance}
\end{figure*}

To better understand the role of the threshold parameter $\tau$ in the pseudoinverse square-root, we performed a threshold sweep on the ill-conditioned ensemble. The numerical results are listed in Table~\ref{tab:exp4_threshold_sweep}, while the retained rank and success probability are shown in Fig.~\ref{fig:threshold_rank} and Fig.~\ref{fig:threshold_success}, respectively. For very small thresholds, $\tau=10^{-12}$ and $\tau=10^{-11}$, all four eigen-components are retained, and the success probability is $0.597538$, with a negligible trace gap of approximately $3.3\times10^{-7}$. Once the threshold increases to $10^{-10}$ and above, the smallest unstable eigenvalue is discarded, reducing the retained rank from $4$ to $3$. Importantly, the success probability remains essentially unchanged at $0.597535$ over a broad interval $10^{-10}\le \tau \le 10^{-2}$. This plateau indicates that the smallest eigenvalue contributes negligibly to discrimination performance while significantly affecting conditioning. Hence, thresholding stabilizes the construction without sacrificing practical accuracy. However, when the threshold is increased to $\tau=10^{-1}$, the retained rank drops further from $3$ to $2$, and the success probability decreases sharply to $0.397975$. This shows that overly aggressive thresholding removes informative spectral components and degrades discrimination performance. Therefore, the threshold should be chosen large enough to suppress unstable near-zero eigenvalues but small enough to preserve the informative support of the ensemble operator.

\begin{table}[t]
\centering
\caption{Threshold sweep results for the pseudoinverse square-root.}
\label{tab:exp4_threshold_sweep}
\begin{tabular}{cccc}
\hline\hline
$\boldsymbol{\tau}$ & \textbf{Rank} & \textbf{Success} & \textbf{Trace Gap} \\
\hline
$1.0\times10^{-12}$ & 4 & 0.597538 & $3.32\times10^{-7}$ \\
$1.0\times10^{-11}$ & 4 & 0.597538 & $3.32\times10^{-7}$ \\
$1.0\times10^{-10}$ & 3 & 0.597535 & 1.000000 \\
$1.0\times10^{-9}$  & 3 & 0.597535 & 1.000000 \\
$1.0\times10^{-8}$  & 3 & 0.597535 & 1.000000 \\
$1.0\times10^{-7}$  & 3 & 0.597535 & 1.000000 \\
$1.0\times10^{-6}$  & 3 & 0.597535 & 1.000000 \\
$1.0\times10^{-5}$  & 3 & 0.597535 & 1.000000 \\
$1.0\times10^{-4}$  & 3 & 0.597535 & 1.000000 \\
$1.0\times10^{-3}$  & 3 & 0.597535 & 1.000000 \\
$1.0\times10^{-2}$  & 3 & 0.597535 & 1.000000 \\
$1.0\times10^{-1}$  & 2 & 0.397975 & 2.000000 \\
\hline
\end{tabular}
\end{table}

\subsubsection{Robustness under depolarizing noise}
To assess the robustness of the proposed method under state corruption, we considered a binary qubit ensemble subjected to depolarizing noise of strength $\lambda \in [0,0.8]$. The numerical values are listed in Table~\ref{tab:exp5_noise_robustness}, and the trend is illustrated in Fig.~\ref{fig:noise_robustness}. At zero noise, the success probability is $0.629410$. As the depolarizing strength increases, the success probability decreases monotonically, reaching $0.501365$ at $\lambda=0.8$. The decrease is initially steep, falling to $0.576057$ already at $\lambda=0.05$, and then gradually saturates near the random-guessing limit of $0.5$, which is expected for highly mixed binary states. This behavior is physically consistent: depolarization reduces distinguishability by shrinking the Bloch vectors toward the maximally mixed state. Importantly, the pseudoinverse-based PGM remains stable throughout the entire noise range and produces sensible discrimination probabilities without numerical breakdown.

\begin{table}[htbp]
\centering
\caption{Noise robustness of the pseudoinverse-based PGM under depolarizing noise.}
\label{tab:exp5_noise_robustness}
\begin{tabular}{lcccccc}
\hline\hline
\multicolumn{7}{c}{$\mathbf{Binary\ qubit\ ensemble\ under\ depolarizing\ noise}$} \\
\hline
\textbf{Noise strength $\lambda$} & \textbf{Success probability} & & & & & \\
\hline
0.000000 & 0.629410 & & & & & \\
0.050000 & 0.576057 & & & & & \\
0.100000 & 0.554894 & & & & & \\
0.150000 & 0.542390 & & & & & \\
0.200000 & 0.533772 & & & & & \\
0.250000 & 0.527331 & & & & & \\
0.300000 & 0.522276 & & & & & \\
0.350000 & 0.518181 & & & & & \\
0.400000 & 0.514796 & & & & & \\
0.450000 & 0.511959 & & & & & \\
0.500000 & 0.509563 & & & & & \\
0.550000 & 0.507531 & & & & & \\
0.600000 & 0.505810 & & & & & \\
0.650000 & 0.504360 & & & & & \\
0.700000 & 0.503150 & & & & & \\
0.750000 & 0.502157 & & & & & \\
0.800000 & 0.501365 & & & & & \\
\hline
\end{tabular}
\end{table}

\subsubsection{Scaling with the number of classes}
We also examined how the pseudoinverse-based PGM scales as the number of classes increases. The results are reported in Table~\ref{tab:exp6_class_scaling} and visualized in Fig.~\ref{fig:class_scaling}. For two classes, the success probability is $1.000000$ in the chosen configuration. As the number of classes grows, the success probability decreases monotonically: $0.652369$ for $K=3$, $0.481036$ for $K=4$, and $0.233599$ for $K=8$. This decrease is expected because increasing the number of classes makes the discrimination problem more crowded in Hilbert space and increases the overlap among candidate states. Nevertheless, the trend remains smooth and well behaved, indicating that the proposed construction scales consistently across multi-class settings. These results suggest that the pseudoinverse-based PGM can serve as a practical discrimination primitive even when the classification problem becomes increasingly complex.

\begin{table}[htbp]
\centering
\caption{Scaling of PGM success probability with the number of classes.}
\label{tab:exp6_class_scaling}
\begin{tabular}{lcccccc}
\hline\hline
\multicolumn{7}{c}{$\mathbf{PGM\ scaling\ with\ number\ of\ classes}$} \\
\hline
\textbf{Number of classes} & \textbf{Success probability} & & & & & \\
\hline
2 & 1.000000 & & & & & \\
3 & 0.652369 & & & & & \\
4 & 0.481036 & & & & & \\
5 & 0.380550 & & & & & \\
6 & 0.314654 & & & & & \\
7 & 0.268153 & & & & & \\
8 & 0.233599 & & & & & \\
\hline
\end{tabular}
\end{table}

\subsubsection{Consistency between direct POVM and Kraus-based construction}
Finally, we numerically verified that the direct PGM POVM construction and the Kraus-operator-based realization are equivalent. The Frobenius norm differences between corresponding POVM elements are shown in Table~\ref{tab:exp7_direct_vs_kraus}. For the two POVM elements considered, the discrepancies are on the order of $10^{-16}$:
\begin{equation}
\|M_0^{(\mathrm{direct})} - M_0^{(\mathrm{Kraus})}\|_F = 7.850462\times10^{-17},
\end{equation}
\begin{equation}
\|M_1^{(\mathrm{direct})} - M_1^{(\mathrm{Kraus})}\|_F = 3.466670\times10^{-16}.
\end{equation}
These values are at the level of machine precision and therefore confirm the numerical equivalence of the two formulations. For completeness, the explicitly reconstructed POVM elements for the small binary case are
\begin{equation}
M_0 =
\begin{bmatrix}
0.6913 & -0.4619 \\
-0.4619 & 0.3087
\end{bmatrix},
\qquad
M_1 =
\begin{bmatrix}
0.3087 & 0.4619 \\
0.4619 & 0.6913
\end{bmatrix},
\end{equation}
with
\begin{equation}
M_0 + M_1 =
\begin{bmatrix}
1 & 0 \\
0 & 1
\end{bmatrix},
\end{equation}
confirming exact completeness for the full-rank case.

\begin{table}[htbp]
\centering
\caption{Comparison between direct PGM POVM construction and Kraus-based construction.}
\label{tab:exp7_direct_vs_kraus}
\begin{tabular}{lcccccc}
\hline\hline
\multicolumn{7}{c}{$\mathbf{Direct\ POVM\ vs.\ Kraus\text{-}based\ construction}$} \\
\hline
\textbf{POVM Element} & \textbf{$\|$Direct $-$ Kraus$\|_F$} & & & & & \\
\hline
$M_0$ & $7.850462\times10^{-17}$ & & & & & \\
$M_1$ & $3.466670\times10^{-16}$ & & & & & \\
\hline
\end{tabular}
\end{table}

\subsection{Circuit-Level Realization on the Real Lupus Dataset}
\label{sec:lupus_circuit_results}

To validate the proposed framework beyond matrix-level simulations, we performed an end-to-end circuit-level implementation on a real binary Lupus dataset. The dataset contains $590$ samples, consisting of $311$ Lupus and $279$ non-Lupus images. Each image was mapped to a two-dimensional feature vector using grayscale mean and standard deviation, followed by normalization. These two features were amplitude-encoded into a single-qubit feature state on register $B$. The dataset was split into $442$ training samples and $148$ test samples, using stratified sampling. From the training data, the class-conditioned density operators were estimated as
\begin{eqnarray}
\sigma_0 &=&
\begin{bmatrix}
0.52989794 & 0.39769796\\
0.39769796 & 0.47010206
\end{bmatrix}\nonumber\\
\sigma_1 &=&
\begin{bmatrix}
0.43245257 & 0.37614968\\
0.37614968 & 0.56754743
\end{bmatrix}.
\end{eqnarray}
Using the empirical class priors $p_0=0.4728506787$ and $p_1=0.5271493213$, the marginal ensemble operator on the feature register was
\begin{equation}
\sigma_B =
\begin{bmatrix}
0.47852968 & 0.38633880\\
0.38633880 & 0.52147032
\end{bmatrix},
\end{equation}
and the joint class-feature operator was
\begin{equation}
\sigma_{XB} =
\resizebox{0.88\columnwidth}{!}{$
\begin{bmatrix}
0.25056260 & 0.18805175 & 0 & 0\\
0.18805175 & 0.22228808 & 0 & 0\\
0 & 0 & 0.22796708 & 0.19828705\\
0 & 0 & 0.19828705 & 0.29918224
\end{bmatrix}
$}.
\end{equation}

The circuit implementation follows the constructive PGM specialization described in Section~\ref{methodology}. First, purified-access block-encodings of $\sigma_B$ and $\sigma_{XB}$ were constructed. The extracted zero-ancilla blocks reproduce the target operators with numerical errors
\begin{equation}
\left\|V_{\sigma_B}-\sigma_B\right\|_F
=
7.72\times 10^{-16},
\end{equation}
and
\begin{equation}
\left\|V_{\sigma_{XB}}-\sigma_{XB}\right\|_F
=
2.73\times 10^{-15}.
\end{equation}
These values confirm that the purified block-encoding stage is exact up to machine precision. Next, the inverse-square-root transformation $\sigma_B^{+1/2}$ was implemented through a QSVT-style spectral transformation. The resulting approximation error was
\begin{equation}
\left\|T_B^{(\mathrm{QSVT})}
-
\sigma_B^{+1/2}/\alpha_B
\right\|_F
=
4.33\times 10^{-5},
\end{equation}
with operator-norm error $4.33\times 10^{-5}$. The $\sigma_{XB}^{1/2}$ transformation was implemented exactly in the present small-dimensional circuit realization, giving zero numerical error for this transformed block. The transformed blocks were then combined to construct the intermediate unitary $\widetilde{W}$, followed by the one-step OAA procedure. The final class probabilities generated by the circuit were compared against the theoretical PGM probabilities on $20$ held-out test samples. The results are summarized in Table~\ref{tab:lupus_circuit_validation}. The raw postselected circuit probabilities agree with the theoretical PGM probabilities with mean TVD on the order of $10^{-6}$. After OAA, the output probabilities remain unchanged within numerical precision, while the mean success probability increases from approximately $0.0565$ to $0.4350$.

\begin{table}[t]
\centering
\caption{End-to-end circuit-level validation of the proposed PGM realization on the real Lupus dataset.}
\label{tab:lupus_circuit_validation}
\begin{tabular}{lcc}
\hline\hline
\textbf{Quantity} & \textbf{Raw circuit} & \textbf{After OAA} \\
\hline
Number of evaluated test samples & 20 & 20 \\
Theory accuracy on evaluated samples & 0.65 & 0.65 \\
Circuit accuracy & 0.65 & 0.65 \\
Mean L1 distance from theory & $2\times 10^{-6}$ & $2\times 10^{-6}$ \\
Mean L2 distance from theory & $2\times 10^{-6}$ & $1\times 10^{-6}$ \\
Mean TVD from theory & $1\times 10^{-6}$ & $1\times 10^{-6}$ \\
Mean success probability & 0.056532 & 0.434977 \\
\hline
\end{tabular}
\end{table}

These results are significant for two reasons. First, they demonstrate that the proposed constructive circuit reproduces the theoretical PGM statistics with essentially machine-precision accuracy. Thus, the implementation is not merely a matrix-level simulation of PGM, but an explicit circuit-level realization of the measurement statistics. Second, the OAA step increases the success probability by a factor of approximately $7.7$ while preserving the class probabilities. This confirms that the amplification procedure improves the practical postselection rate without distorting the PGM decision rule. It should be emphasized that the purpose of this experiment is not to claim state-of-the-art classical classification accuracy on the Lupus dataset. Rather, the central objective is to verify that the proposed block-encoded PGM construction faithfully realizes the theoretical measurement on real data. The observed accuracy of $0.65$ reflects the limited two-feature amplitude encoding used in this proof-of-concept implementation. The key result is the near-perfect agreement between theoretical PGM probabilities and circuit-generated probabilities, together with a substantial OAA-based improvement in success probability.

\begin{table*}[t]
\centering
\caption{Summary of PGM experiments comparing inverse and pseudoinverse constructions.}
\label{tab:pgm_main_summary}
\begin{tabular}{lcccccc}
\hline\hline
\textbf{Case} & \textbf{dim} & \textbf{$K$} & \textbf{rank} & \textbf{eig range} & \textbf{Inv} & \textbf{Pseudo (Success, Gap)} \\
\hline
Well-cond. (2D) & 2 & 2 & 2 & $[0.038,\;0.962]$ & OK & $(0.691,\;0.000)$ \\
Rank-def. (4D)  & 4 & 3 & 2 & $[0,\;0.700]$     & Fail & $(0.639,\;2.000)$ \\
Ill-cond. (4D)  & 4 & 4 & 3 & $[0,\;0.797]$     & Fail & $(0.598,\;1.000)$ \\
\hline
\end{tabular}
\end{table*}
A consolidated summary of the three principal structural experiments is provided in Table~\ref{tab:pgm_main_summary}. The results reveal three distinct regimes. In the well-conditioned full-rank case, the inverse-based and pseudoinverse-based constructions coincide exactly, confirming consistency with the standard PGM formulation. In the rank-deficient case, the inverse-based construction fails completely, whereas the pseudoinverse remains well defined and yields a meaningful success probability. In the ill-conditioned regime, the inverse-based approach becomes numerically unstable, while the pseudoinverse-based method maintains robustness and stability. These observations demonstrate that the pseudoinverse is not merely a mathematical convenience but a necessary component for practical PGM realization in singular and near-singular settings. The experiments further establish four key conclusions: (i) the pseudoinverse-based PGM recovers the standard PGM in well-conditioned full-rank scenarios; (ii) it remains operationally valid in rank-deficient and ill-conditioned regimes where the ordinary inverse fails; (iii) thresholding serves as an effective mechanism to regularize unstable eigenvalues without compromising discrimination accuracy across a wide range of tolerances; and (iv) the Kraus-based realization is numerically indistinguishable from the direct POVM construction. These results support the proposed framework as a robust computational pathway toward practical and circuit-level realization of PGM for realistic quantum ensembles.

\subsection{Structural Regimes and Robustness Analysis}
\begin{figure*}[t]
\centering

\begin{subfigure}[t]{0.49\textwidth}
    \centering
    \includegraphics[width=\linewidth]{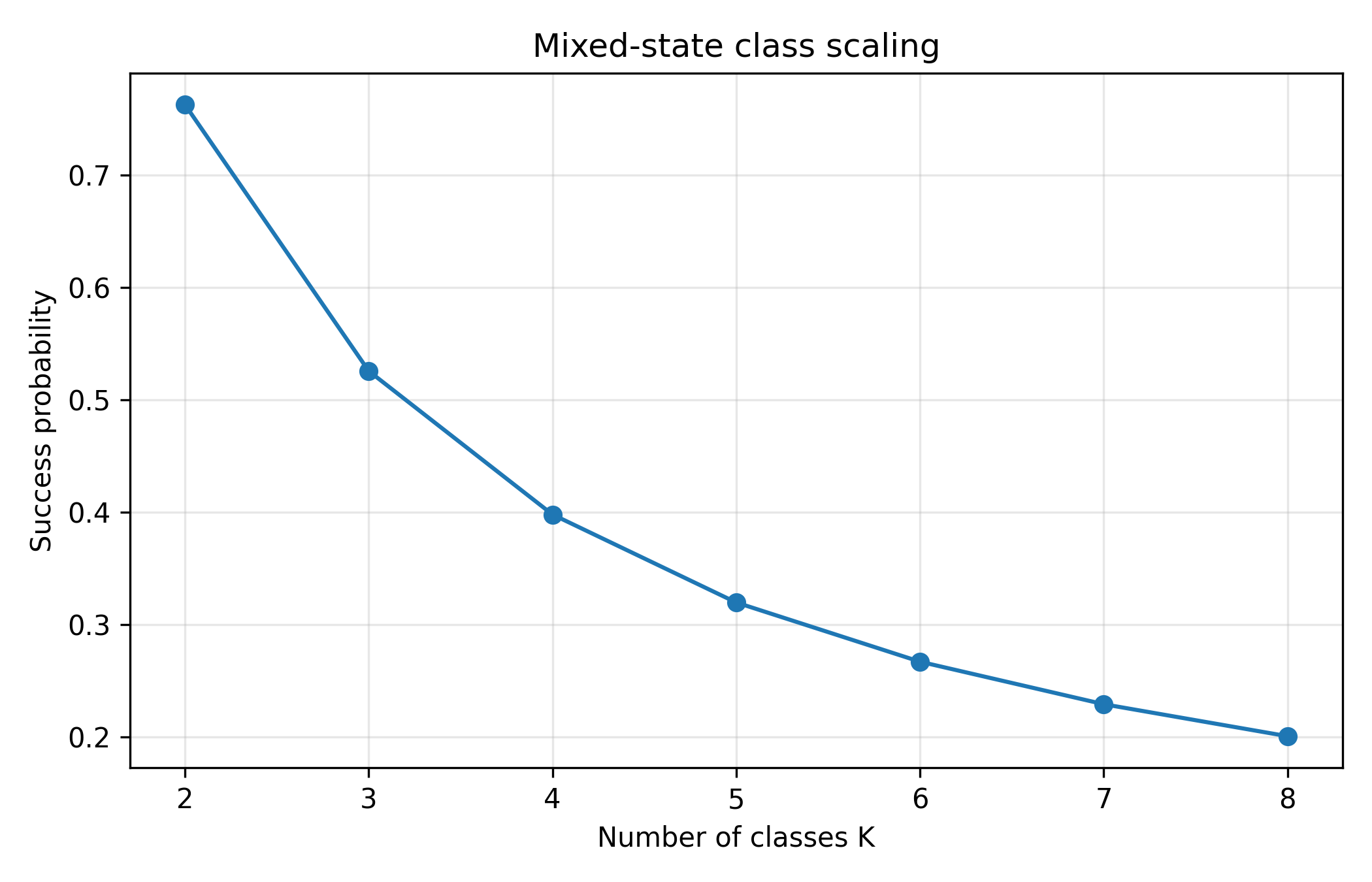}
    \caption{Success probability vs number of classes $K$.}
    \label{fig:class_scaling_psucc}
\end{subfigure}
\hfill
\begin{subfigure}[t]{0.49\textwidth}
    \centering
    \includegraphics[width=\linewidth]{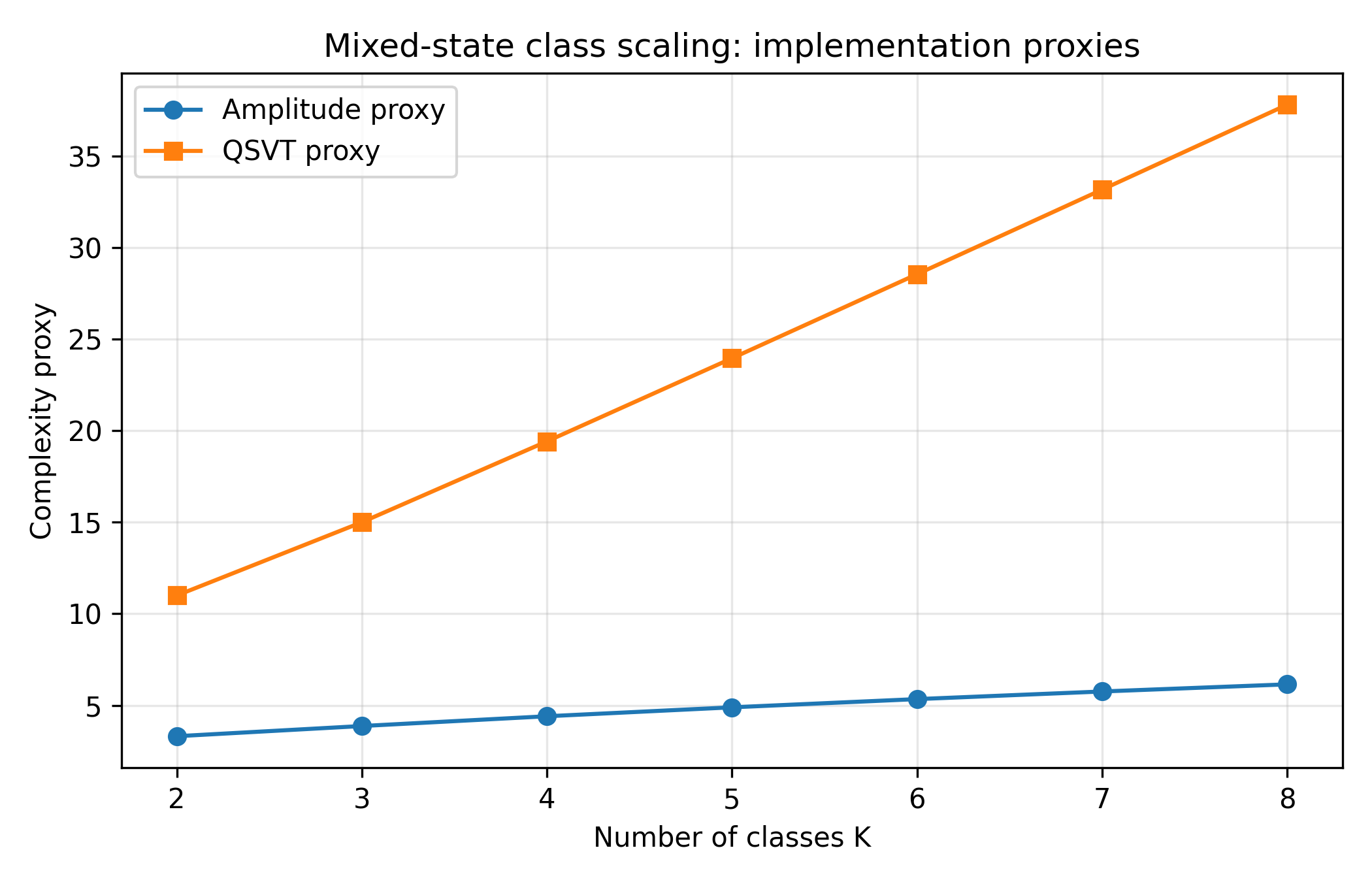}
    \caption{Implementation complexity proxies vs $K$.}
    \label{fig:class_scaling_proxy}
\end{subfigure}

\vspace{0.5em}

\begin{subfigure}[t]{0.49\textwidth}
    \centering
    \includegraphics[width=\linewidth]{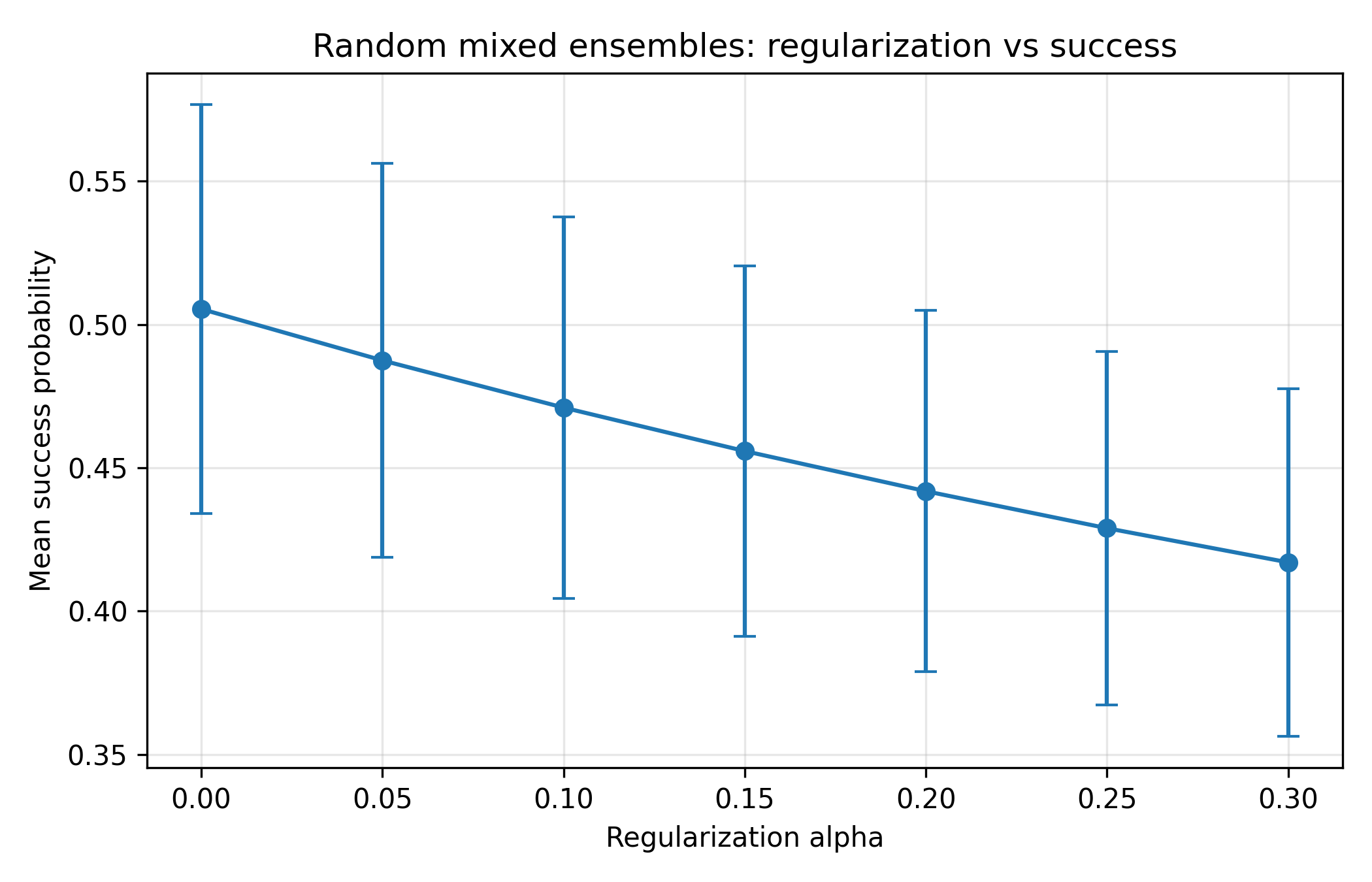}
    \caption{Mean success probability vs regularization strength $\alpha$.}
    \label{fig:regularization_psucc}
\end{subfigure}
\hfill
\begin{subfigure}[t]{0.49\textwidth}
    \centering
    \includegraphics[width=\linewidth]{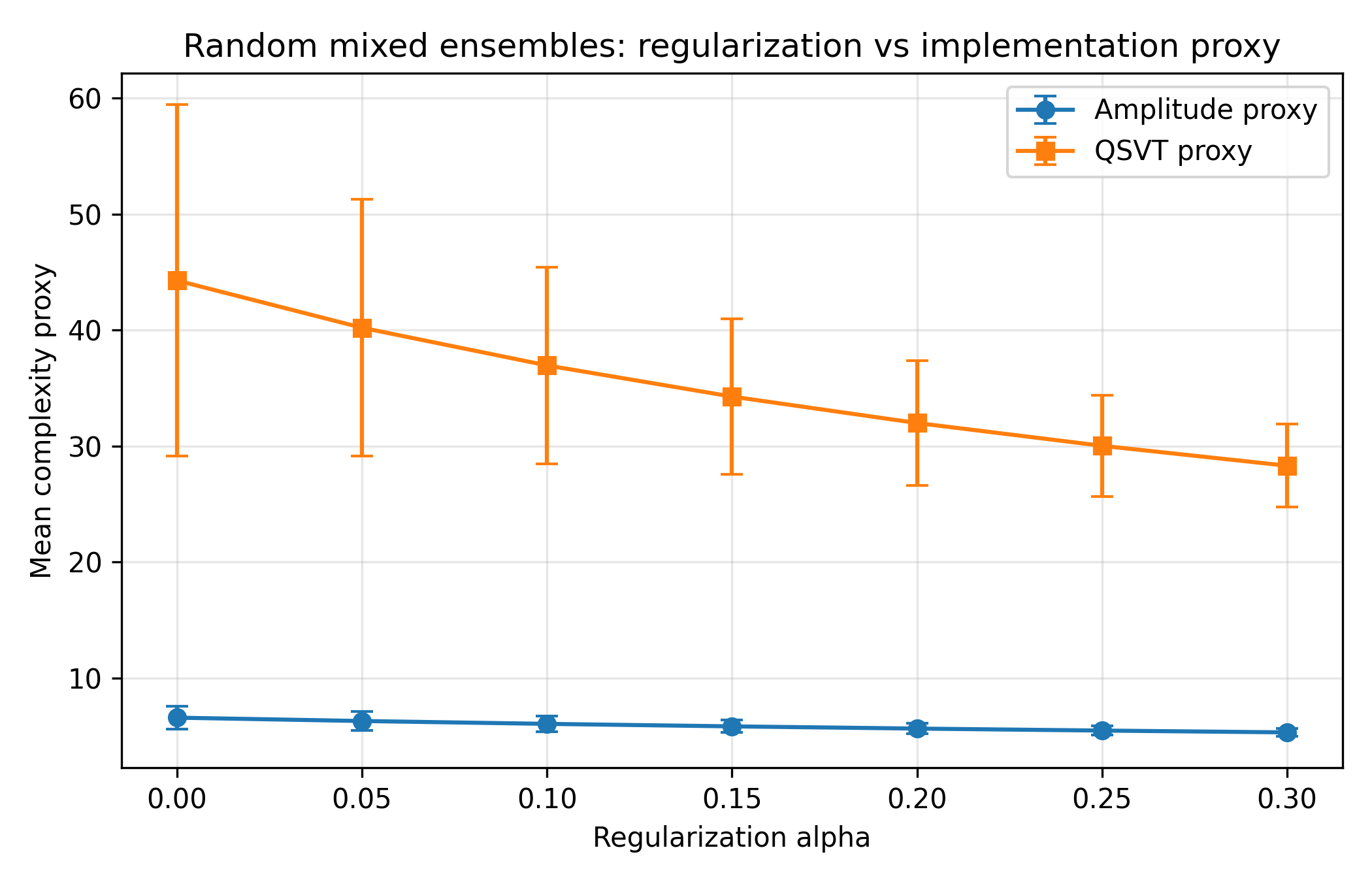}
    \caption{Implementation complexity proxies vs $\alpha$.}
    \label{fig:regularization_proxy}
\end{subfigure}

\caption{
Scaling and regularization behavior of the pseudoinverse-based PGM.
(a) Success probability decreases with increasing number of classes $K$, reflecting reduced distinguishability among quantum states.
(b) Implementation complexity proxies increase with $K$, indicating higher resource requirements.
(c) Mean success probability decreases with regularization strength $\alpha$, as stronger mixing reduces state separability.
(d) Implementation complexity proxies decrease with $\alpha$, showing improved spectral conditioning and reduced implementation cost.
Error bars (where applicable) denote standard deviation over random mixed ensembles.
}
\label{fig:pgm_scaling_regularization}
\end{figure*}

Fig.~\ref{fig:pgm_scaling_regularization} illustrates the trade-off between discrimination performance and implementation complexity in the pseudoinverse-based PGM framework for mixed-state ensembles. As the number of classes $K$ increases, the quantum states become more densely distributed in Hilbert space, leading to increased overlap and a consequent reduction in the success probability, as shown in Fig.~\ref{fig:class_scaling_psucc}. Simultaneously, the complexity proxies, which are inspired by condition-number-dependent costs in QSVT and amplitude amplification-based implementations, grow monotonically with $K$, as shown in Fig.~\ref{fig:class_scaling_proxy}. This demonstrates that multi-class quantum state discrimination becomes both statistically harder and computationally more demanding, highlighting a fundamental trade-off between accuracy and implementation cost in practical realizations of PGM. Fig.~\ref{fig:pgm_scaling_regularization} shows how isotropic regularization affects both the discrimination performance and the implementation-oriented complexity of the pseudoinverse-based PGM in random mixed-state ensembles. As shown in Fig.~\ref{fig:regularization_psucc}, the mean success probability decreases smoothly with increasing regularization strength $\alpha$, which is expected because mixing each state with the maximally mixed state reduces the distinguishability between classes. At the same time, Fig.~\ref{fig:regularization_proxy} shows that both the amplitude-based and QSVT-inspired complexity proxies decrease monotonically as $\alpha$ increases, indicating that regularization improves the spectral profile of the ensemble operator and makes inverse-like transformations more stable and less demanding. Thus, the figure highlights a clear trade-off: moderate regularization can substantially reduce implementation burden while causing only a gradual degradation in discrimination performance. This behavior supports the role of regularization as a practically useful mechanism for stabilizing support-aware PGM constructions in mixed and potentially ill-conditioned settings.

\begin{figure*}[t]
\centering

\begin{subfigure}[t]{0.49\textwidth}
    \centering
    \includegraphics[width=\linewidth]{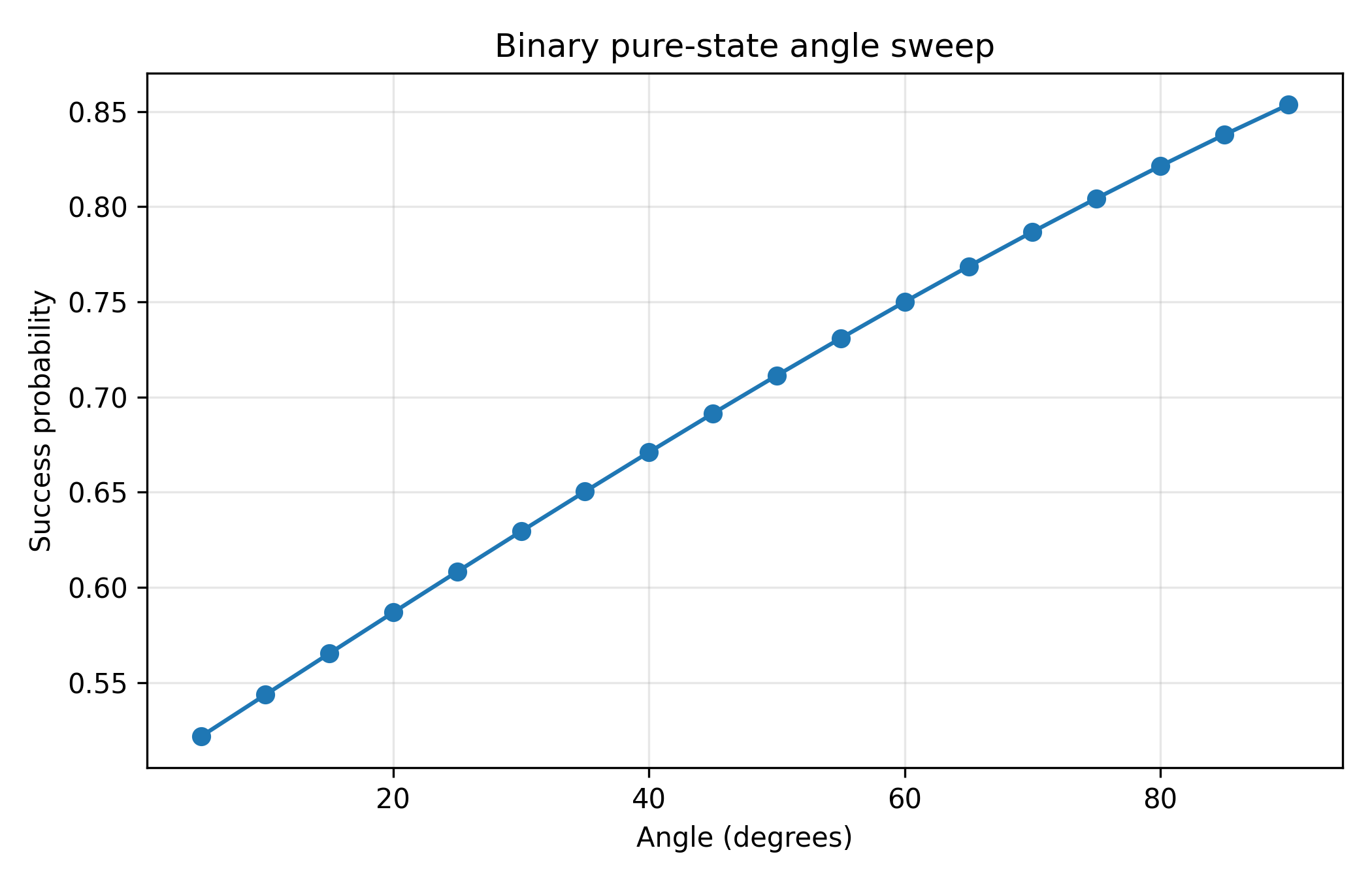}
    \caption{Success probability vs state separation angle.}
    \label{fig:angle_psucc}
\end{subfigure}
\hfill
\begin{subfigure}[t]{0.49\textwidth}
    \centering
    \includegraphics[width=\linewidth]{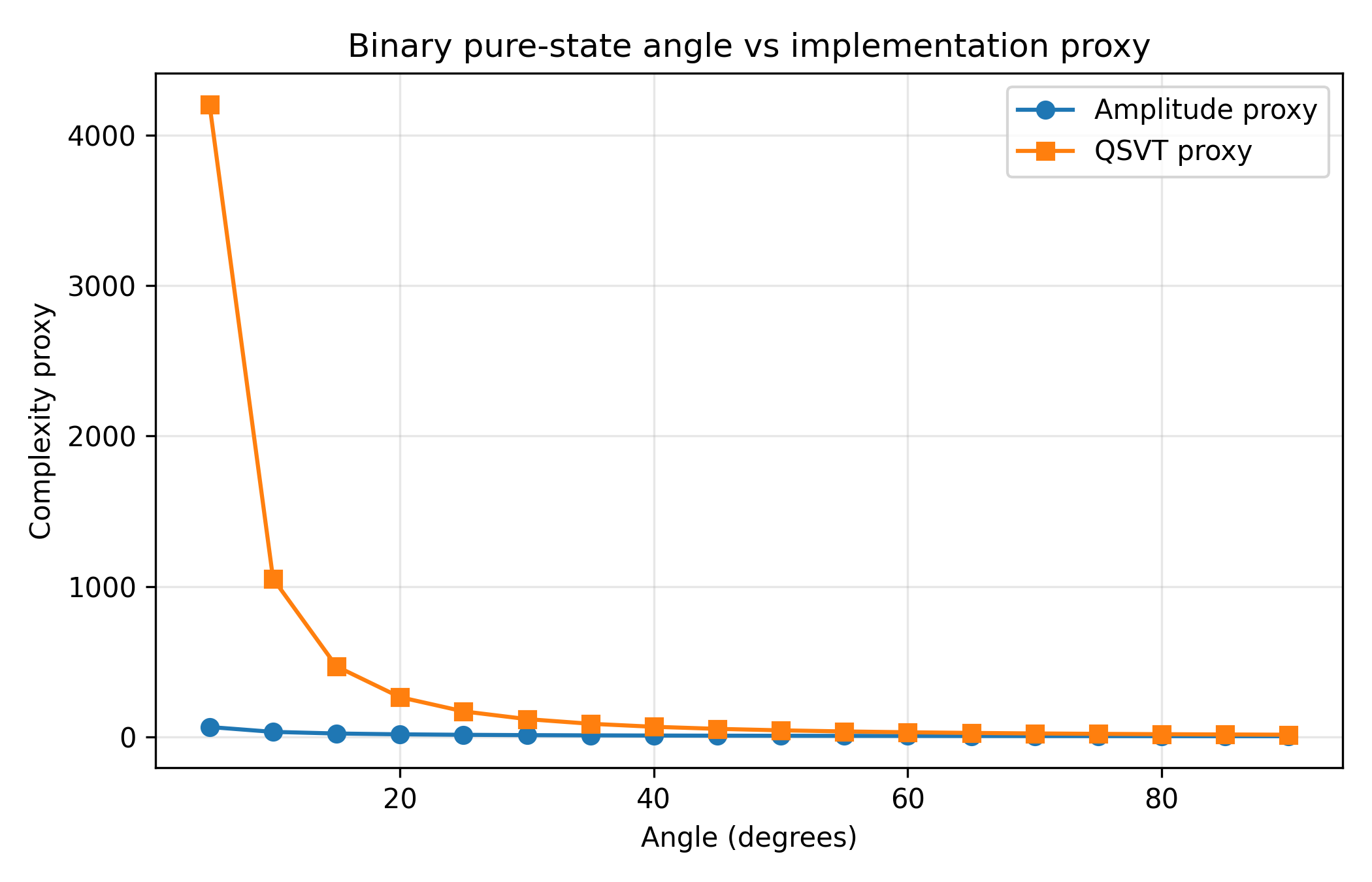}
    \caption{Implementation complexity proxies vs state separation angle.}
    \label{fig:angle_proxy}
\end{subfigure}

\vspace{0.5em}

\begin{subfigure}[t]{0.49\textwidth}
    \centering
    \includegraphics[width=\linewidth]{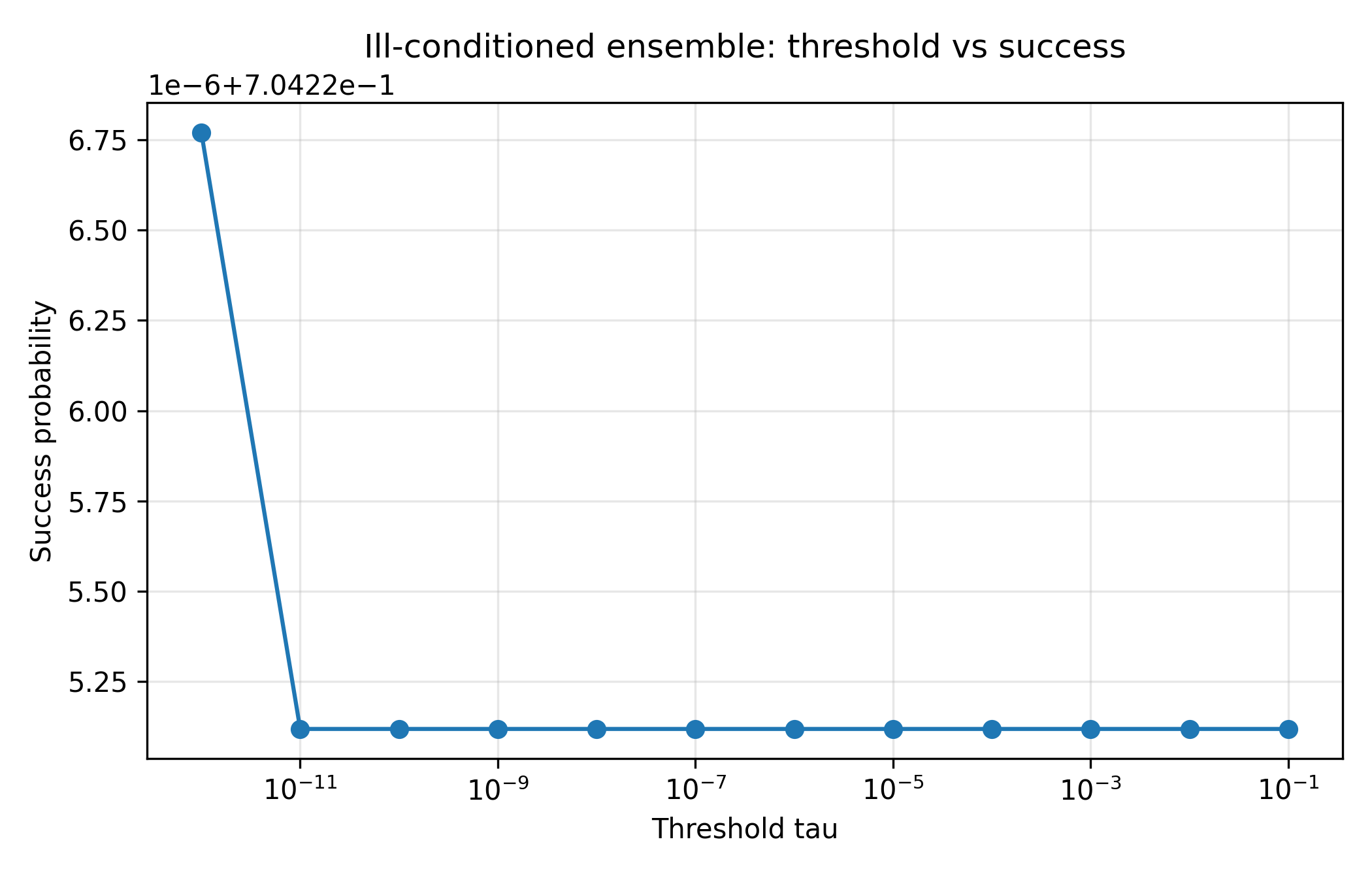}
    \caption{Success probability vs threshold $\tau$.}
    \label{fig:tau_psucc}
\end{subfigure}
\hfill
\begin{subfigure}[t]{0.49\textwidth}
    \centering
    \includegraphics[width=\linewidth]{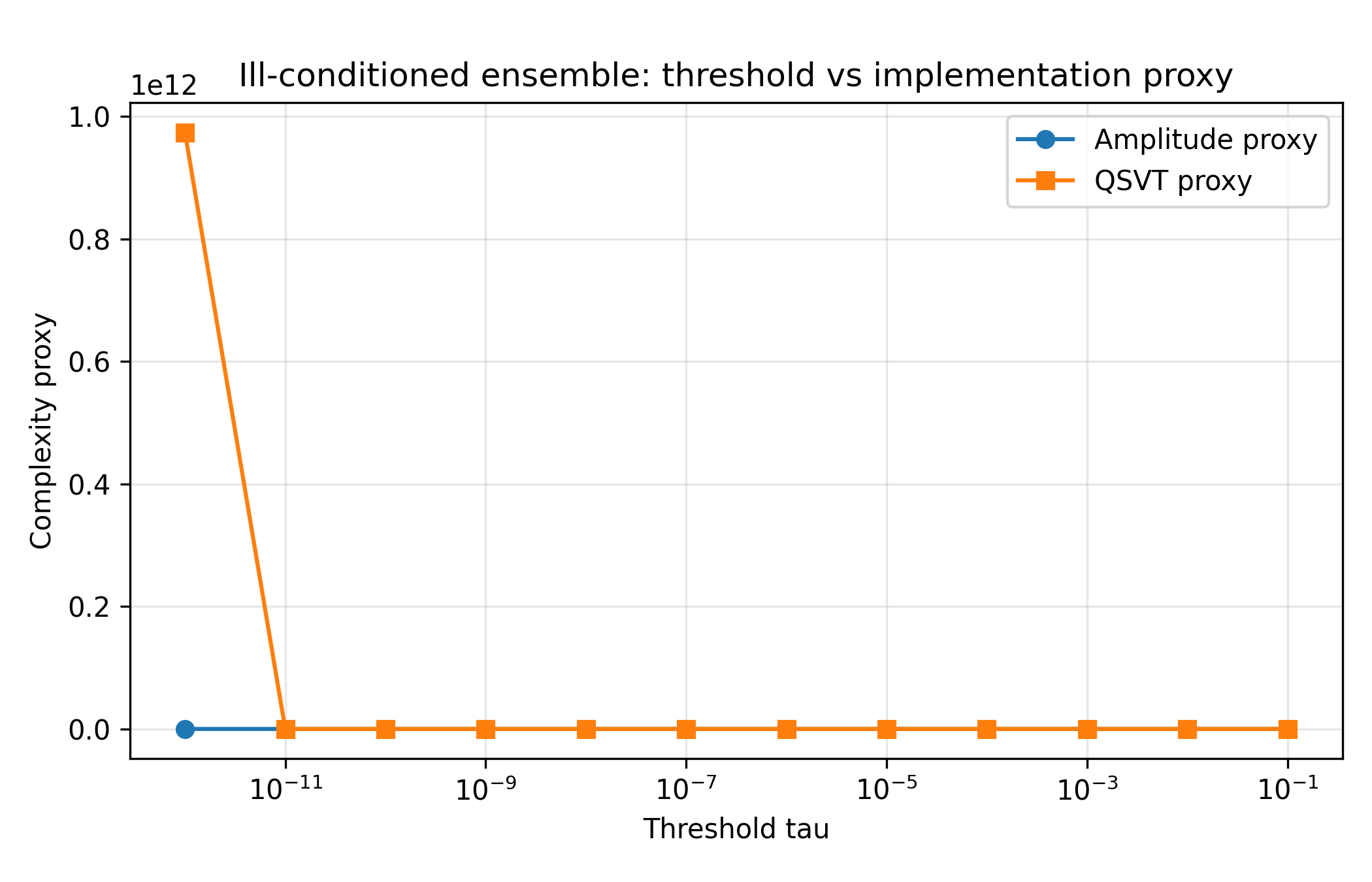}
    \caption{Implementation complexity proxies vs $\tau$.}
    \label{fig:tau_proxy}
\end{subfigure}

\caption{
Performance and conditioning behavior of the pseudoinverse-based PGM under geometric separation and spectral regularization.
(a) Success probability increases with the angle between quantum states, reflecting improved distinguishability.
(b) Implementation complexity proxies decrease with increasing angle, indicating better conditioning of the ensemble operator.
(c) Success probability remains stable across a wide range of threshold values $\tau$, demonstrating robustness to spectral truncation.
(d) Implementation complexity proxies decrease significantly with increasing $\tau$, showing that removing small eigenvalue components improves conditioning and reduces computational cost.
}
\label{fig:pgm_angle_threshold}
\end{figure*}

Fig.~\ref{fig:pgm_angle_threshold} illustrates the relationship between state distinguishability and implementation complexity in the binary pure-state setting. As shown in Fig.~\ref{fig:angle_psucc}, the success probability increases monotonically with the separation angle between the two quantum states, since larger angles correspond to reduced overlap and hence easier discrimination. In contrast, Fig.~\ref{fig:angle_proxy} shows that both amplitude-based and QSVT-inspired complexity proxies decrease rapidly as the angle increases, particularly in the small-angle regime where the states are nearly indistinguishable and the ensemble operator becomes highly ill-conditioned. This inverse relationship highlights a fundamental principle: quantum state discrimination is most computationally demanding precisely when the states are hardest to distinguish. As the angle grows, the spectral conditioning improves, leading to more stable pseudoinverse operations and significantly reduced implementation cost. Fig.~\ref{fig:pgm_angle_threshold} demonstrates the effect of threshold regularization on the pseudoinverse-based PGM in an ill-conditioned ensemble. As shown in Fig.~\ref{fig:tau_psucc}, the success probability remains essentially unchanged once the threshold $\tau$ exceeds the scale of numerical noise, indicating that small eigenvalue components do not contribute significantly to discrimination performance. In contrast, Fig.~\ref{fig:tau_proxy} shows a sharp reduction in both amplitude-based and QSVT-inspired complexity proxies as $\tau$ increases, particularly in the transition from extremely small thresholds to moderate values. This reflects the removal of poorly conditioned spectral components that would otherwise dominate the cost of inverse-like transformations. The results highlight a key advantage of threshold regularization: it enables substantial reductions in implementation complexity with negligible impact on performance, making it a practical and effective strategy for stabilizing PGM constructions in ill-conditioned quantum systems.

\begin{table*}[t]
\centering
\caption{Binary pure-state angle sweep results for the pseudoinverse-based PGM.}
\label{tab:angle_sweep_exact}
\begin{tabular}{c c c c c c c}
\hline\hline
Angle ($^\circ$) & $P_{\mathrm{succ}}$ & $\lambda_{\min}^{+}(\sigma_B)$ & $\kappa(\sigma_B)$ & Amp. Proxy & QSVT Proxy & $\|M^{(\mathrm{dir})}-M^{(\mathrm{Kraus})}\|_{F,\max}$ \\
\hline
 5  & 0.521810 & 0.000476 & 2100.329429 & 64.827917 & 4202.658858 & $9.238897\times10^{-16}$ \\
10  & 0.543578 & 0.001903 &  524.582476 & 32.421674 & 1051.164953 & $0$ \\
15  & 0.565263 & 0.004278 &  232.777626 & 21.623026 &  467.555252 & $2.355139\times10^{-16}$ \\
20  & 0.586824 & 0.007596 &  130.646096 & 16.226281 &  263.292191 & $1.359740\times10^{-16}$ \\
25  & 0.608220 & 0.011852 &   83.373971 & 12.990302 &  168.747941 & $6.030055\times10^{-16}$ \\
30  & 0.629410 & 0.017037 &   57.695481 & 10.834711 &  117.390961 & $5.147892\times10^{-16}$ \\
35  & 0.650353 & 0.023142 &   42.212365 &  9.296490 &   86.424729 & $8.326673\times10^{-16}$ \\
40  & 0.671010 & 0.030154 &   32.163437 &  8.144131 &   66.326875 & $8.196123\times10^{-16}$ \\
45  & 0.691342 & 0.038060 &   25.274142 &  7.249020 &   52.548285 & $2.077037\times10^{-16}$ \\
50  & 0.711309 & 0.046846 &   20.346491 &  6.533987 &   42.692982 & $3.040471\times10^{-16}$ \\
55  & 0.730874 & 0.056495 &   16.700812 &  5.949926 &   35.401624 & $7.216450\times10^{-16}$ \\
60  & 0.750000 & 0.066987 &   13.928203 &  5.464102 &   29.856406 & $4.710277\times10^{-16}$ \\
65  & 0.768650 & 0.078304 &   11.770694 &  5.053849 &   25.541389 & $4.878985\times10^{-16}$ \\
70  & 0.786788 & 0.090424 &   10.059014 &  4.702981 &   22.118027 & $4.088652\times10^{-16}$ \\
75  & 0.804381 & 0.103323 &    8.678356 &  4.399626 &   19.356713 & $2.371437\times10^{-16}$ \\
80  & 0.821394 & 0.116978 &    7.548632 &  4.134884 &   17.097264 & $3.845925\times10^{-16}$ \\
85  & 0.837795 & 0.131361 &    6.612590 &  3.901946 &   15.225181 & $4.227603\times10^{-16}$ \\
90  & 0.853553 & 0.146447 &    5.828427 &  3.695518 &   13.656854 & $2.419675\times10^{-16}$ \\
\hline
\end{tabular}
\end{table*}

Table~\ref{tab:angle_sweep_exact} reports the behavior of the pseudoinverse-based PGM for binary pure-state discrimination as the angular separation between the states increases. The success probability rises monotonically from $0.521810$ at $5^\circ$ to $0.853553$ at $90^\circ$, confirming that larger angular separation improves distinguishability. At the same time, the smallest nonzero eigenvalue of the ensemble operator increases from $0.000476$ to $0.146447$, while the effective condition number decreases sharply from $2100.329429$ to $5.828427$. This directly translates into a large reduction in the implementation-oriented cost proxies: the amplitude proxy drops from $64.827917$ to $3.695518$, and the QSVT proxy drops from $4202.658858$ to $13.656854$. Furthermore, the maximum Frobenius-norm discrepancy between the direct POVM construction and the Kraus-based construction remains at the level of numerical precision throughout, confirming structural consistency of the two realizations. The table shows that binary state discrimination becomes simultaneously more accurate and less computationally demanding as the states become more separated.

\begin{table*}[t]
\centering
\caption{Threshold sweep results for the ill-conditioned ensemble under the pseudoinverse-based PGM.}
\label{tab:threshold_sweep_exact}
\begin{tabular}{c c c c c c c}
\hline\hline
$\tau$ & Rank & $P_{\mathrm{succ}}$ & Trace Gap (full) & Trace Gap (support) & Amp. Proxy & QSVT Proxy \\
\hline
$1.0\times10^{-12}$ & 4 & 0.704227 & $3.885781\times10^{-15}$ & $3.219647\times10^{-15}$ & 986576.572467 & $9.733333\times10^{11}$ \\
$1.0\times10^{-11}$ & 3 & 0.704225 & 1.000000 & $1.110223\times10^{-16}$ & 5.734156 & 32.880540 \\
$1.0\times10^{-10}$ & 3 & 0.704225 & 1.000000 & $1.110223\times10^{-16}$ & 5.734156 & 32.880540 \\
$1.0\times10^{-9}$  & 3 & 0.704225 & 1.000000 & $1.110223\times10^{-16}$ & 5.734156 & 32.880540 \\
$1.0\times10^{-8}$  & 3 & 0.704225 & 1.000000 & $1.110223\times10^{-16}$ & 5.734156 & 32.880540 \\
$1.0\times10^{-7}$  & 3 & 0.704225 & 1.000000 & $1.110223\times10^{-16}$ & 5.734156 & 32.880540 \\
$1.0\times10^{-6}$  & 3 & 0.704225 & 1.000000 & $1.110223\times10^{-16}$ & 5.734156 & 32.880540 \\
$1.0\times10^{-5}$  & 3 & 0.704225 & 1.000000 & $1.110223\times10^{-16}$ & 5.734156 & 32.880540 \\
$1.0\times10^{-4}$  & 3 & 0.704225 & 1.000000 & $1.110223\times10^{-16}$ & 5.734156 & 32.880540 \\
$1.0\times10^{-3}$  & 3 & 0.704225 & 1.000000 & $1.110223\times10^{-16}$ & 5.734156 & 32.880540 \\
$1.0\times10^{-2}$  & 3 & 0.704225 & 1.000000 & $1.110223\times10^{-16}$ & 5.734156 & 32.880540 \\
$1.0\times10^{-1}$  & 3 & 0.704225 & 1.000000 & $1.110223\times10^{-16}$ & 5.734156 & 32.880540 \\
\hline
\end{tabular}
\end{table*}

Table~\ref{tab:threshold_sweep_exact} shows the effect of spectral thresholding on an ill-conditioned ensemble. At the smallest threshold $\tau=10^{-12}$, the effective rank is $4$, the success probability is $0.704227$, and both trace gaps are essentially zero, but the implementation proxies are extremely large, with amplitude proxy $986576.572467$ and QSVT proxy $9.733333\times10^{11}$, revealing severe numerical instability caused by the tiny retained eigenvalue. Once the threshold is increased to $10^{-11}$, the effective rank drops to $3$, the success probability changes only marginally to $0.704225$, and the support trace gap remains at numerical zero, while the proxies collapse to $5.734156$ and $32.880540$, respectively. This behavior remains unchanged for all larger thresholds listed. The full-space trace gap of $1.000000$ in the thresholded regime reflects that the measurement becomes complete on the support of the ensemble operator rather than on the full Hilbert space. Thus, the table provides strong evidence that threshold regularization removes spectrally unstable components at negligible cost to discrimination performance while dramatically reducing implementation burden.

\begin{table}[t]
\centering
\caption{Mixed-state class scaling results for the pseudoinverse-based PGM.}
\label{tab:class_scaling_exact}

\resizebox{\columnwidth}{!}{%
\begin{tabular}{c c c c c c}
\hline\hline
$K$ & $P_{\mathrm{succ}}$ & $\lambda_{\min}^{+}(\sigma_B)$ & $\kappa(\sigma_B)$ & Amp. Proxy & QSVT Proxy \\
\hline
2 & 0.762523 & 0.181802 & 4.500491 & 3.316773 & 11.000982 \\
3 & 0.525521 & 0.200000 & 4.000000 & 3.872983 & 15.000000 \\
4 & 0.397832 & 0.206023 & 3.853819 & 4.406277 & 19.415277 \\
5 & 0.319540 & 0.208754 & 3.790330 & 4.894042 & 23.951650 \\
6 & 0.266841 & 0.210222 & 3.756870 & 5.342399 & 28.541222 \\
7 & 0.229004 & 0.211103 & 3.737033 & 5.758405 & 33.159228 \\
8 & 0.200537 & 0.211672 & 3.724290 & 6.147709 & 37.794321 \\
\hline
\end{tabular}%
}
\end{table}

Table~\ref{tab:class_scaling_exact} summarizes the effect of increasing the number of classes in the mixed-state discrimination setting. The success probability decreases monotonically from $0.762523$ for $K=2$ to $0.200537$ for $K=8$, indicating that discrimination becomes progressively harder as the ensemble becomes more crowded. Interestingly, the smallest nonzero eigenvalue of the ensemble operator increases slightly from $0.181802$ to $0.211672$, and the effective condition number decreases mildly from $4.500491$ to $3.724290$, showing that the ensemble operator itself does not become more ill-conditioned. Nevertheless, the implementation proxies increase steadily with $K$, with amplitude proxy rising from $3.316773$ to $6.147709$ and QSVT proxy from $11.000982$ to $37.794321$. This reflects the fact that implementation cost grows not only through conditioning but also through the increasing size and complexity of the multiclass discrimination problem. Hence, the table highlights a practical trade-off: even when spectral conditioning remains controlled, scaling to more classes still reduces accuracy and increases computational burden.

\begin{table*}[t]
\centering
\caption{Regularization effects on random mixed ensembles under the pseudoinverse-based PGM.}
\label{tab:regularization_exact}
\begin{tabular}{c c c c c c c}
\hline\hline
$\alpha$ & Mean $P_{\mathrm{succ}}$ & Std. $P_{\mathrm{succ}}$ & Mean Amp. Proxy & Std. Amp. Proxy & Mean QSVT Proxy & Std. QSVT Proxy \\
\hline
0.00 & 0.505431 & 0.071251 & 6.577226 & 1.005935 & 44.271807 & 15.150242 \\
0.05 & 0.487494 & 0.068703 & 6.289265 & 0.801929 & 40.197940 & 11.070981 \\
0.10 & 0.471028 & 0.066501 & 6.043017 & 0.656337 & 36.948827 &  8.475315 \\
0.15 & 0.455861 & 0.064619 & 5.827719 & 0.546052 & 34.260481 &  6.677995 \\
0.20 & 0.441866 & 0.063034 & 5.636579 & 0.459160 & 31.981856 &  5.362708 \\
0.25 & 0.428947 & 0.061717 & 5.464932 & 0.388762 & 30.016621 &  4.362128 \\
0.30 & 0.417032 & 0.060644 & 5.309391 & 0.330521 & 28.298874 &  3.578864 \\
\hline
\end{tabular}
\end{table*}

Table~\ref{tab:regularization_exact} presents the effect of isotropic regularization on random mixed-state ensembles. As the regularization strength $\alpha$ increases from $0.00$ to $0.30$, the mean success probability decreases gradually from $0.505431$ to $0.417032$, while its standard deviation decreases from $0.071251$ to $0.060644$. At the same time, the mean amplitude proxy decreases from $6.577226$ to $5.309391$, and the mean QSVT proxy decreases from $44.271807$ to $28.298874$, with their corresponding standard deviations also shrinking substantially. These trends indicate that regularization improves the spectral stability of the ensemble operator and makes the implementation burden more predictable across random instances, albeit at the cost of some reduction in discrimination performance. Therefore, the table quantitatively exhibits a favorable stability-accuracy trade-off: moderate regularization can significantly reduce implementation complexity and variability while causing only a smooth degradation in mean success probability.

\subsection{Sampling Effects and Training Stability Analysis}
\begin{figure*}[t]
\centering

\begin{subfigure}[t]{0.32\textwidth}
    \centering
    \includegraphics[width=\linewidth]{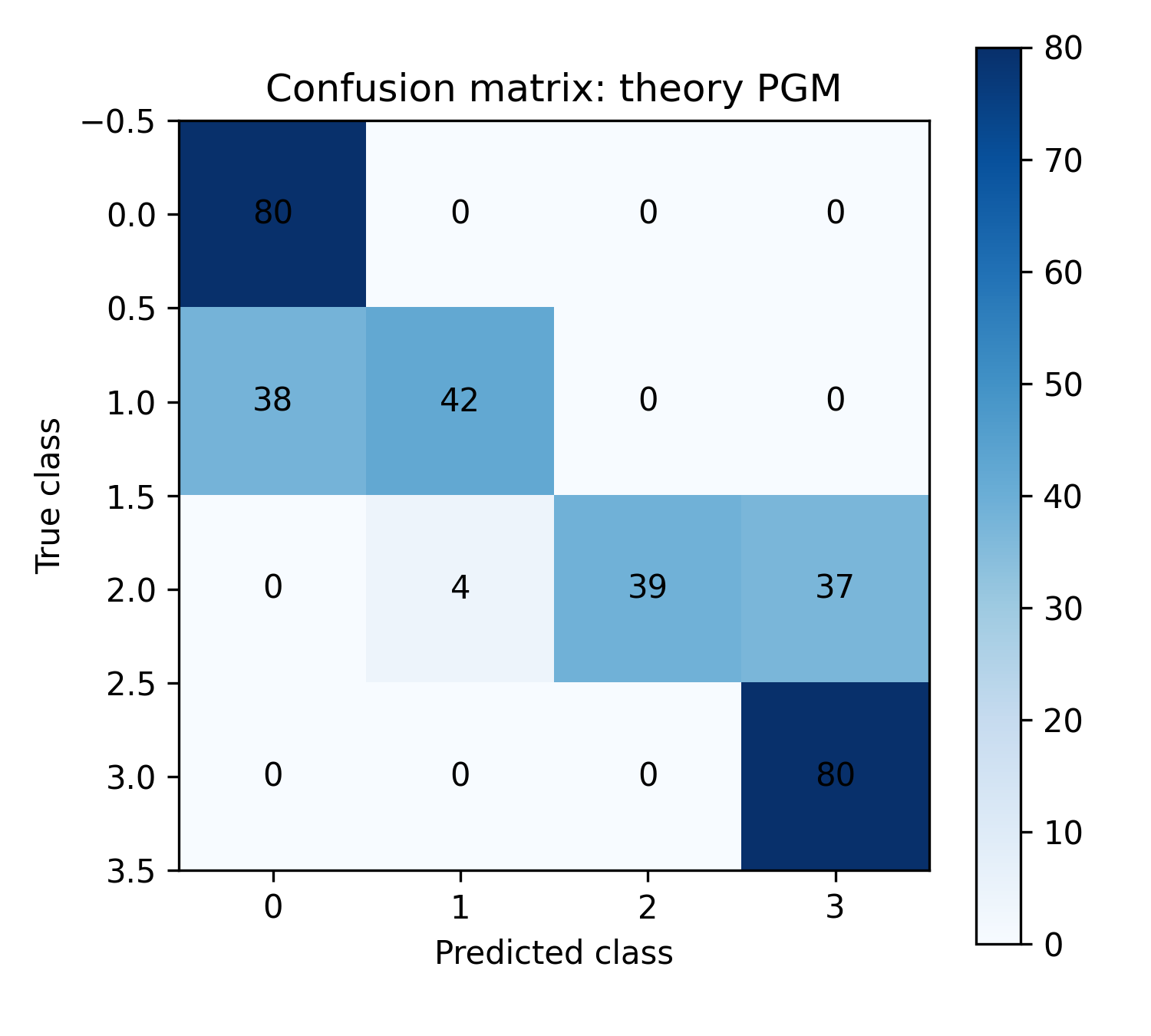}
    \caption{Theory PGM}
    \label{fig:confusion_theory}
\end{subfigure}
\hfill
\begin{subfigure}[t]{0.32\textwidth}
    \centering
    \includegraphics[width=\linewidth]{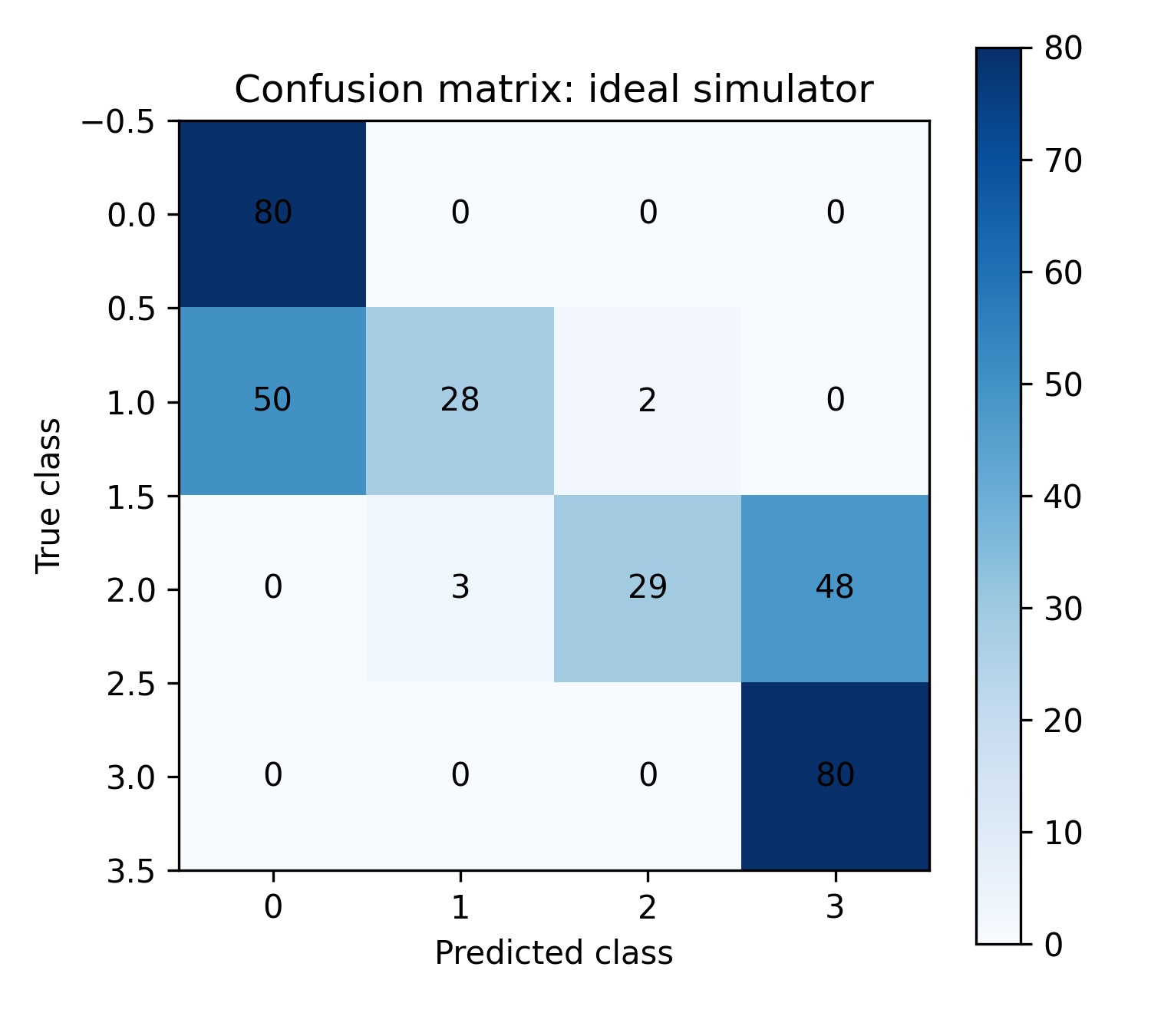}
    \caption{Ideal simulator}
    \label{fig:confusion_ideal}
\end{subfigure}
\hfill
\begin{subfigure}[t]{0.32\textwidth}
    \centering
    \includegraphics[width=\linewidth]{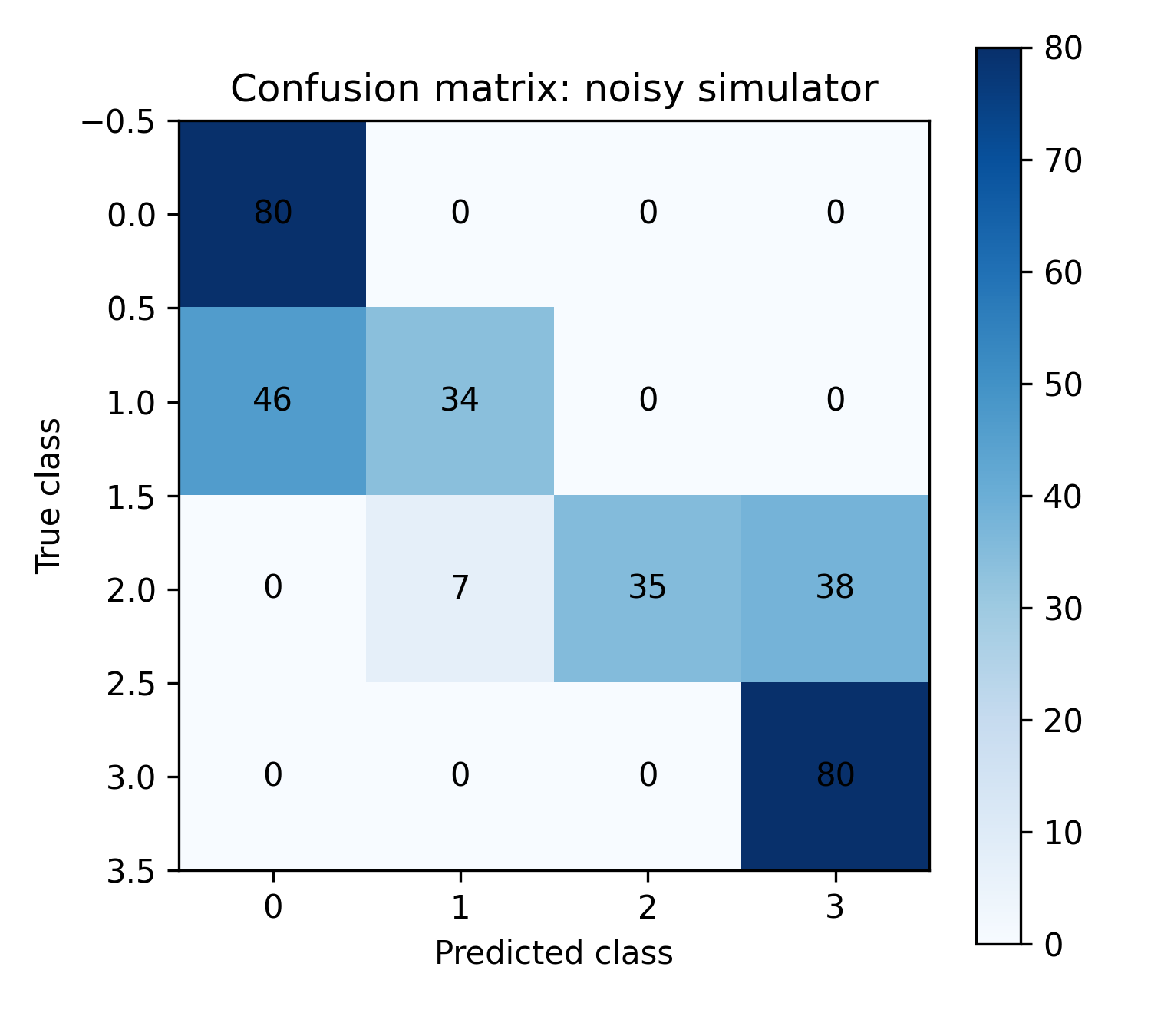}
    \caption{Noisy simulator}
    \label{fig:confusion_noisy}
\end{subfigure}

\caption{
Confusion matrices for the four-class trained PGM classifier.
(a) Theoretical PGM predictions obtained directly from the trained POVM elements.
(b) End-to-end implementation on the ideal quantum simulator using the Naimark-dilation-based circuit realization.
(c) End-to-end implementation on the noisy quantum simulator.
All three panels show strong agreement on classes 0 and 3, while the dominant classification difficulty occurs between the middle classes 1 and 2, where the underlying states are more strongly overlapping.
}
\label{fig:confusion_matrices_full}
\end{figure*}

Fig.~\ref{fig:confusion_matrices_full} presents the confusion matrices of the trained four-class PGM classifier at three levels of analysis: the matrix-level theoretical PGM rule in Fig.~\ref{fig:confusion_theory}, the full circuit-based realization on the ideal simulator in Fig.~\ref{fig:confusion_ideal}, and the same implementation under noise in Fig.~\ref{fig:confusion_noisy}. A first important observation is that classes 0 and 3 are classified almost perfectly across all settings, with all 80 test samples mapped correctly in each case. This indicates that these edge classes are well separated in the chosen quantum feature space and remain robust even after circuit execution and noise. The main classification errors are concentrated in the intermediate classes 1 and 2, which is consistent with the fact that these classes are geometrically closer and therefore harder to discriminate. More specifically, the theoretical PGM in Fig.~\ref{fig:confusion_theory} predicts that class 1 is split mainly between labels 0 and 1, with counts $(38,42,0,0)$, while class 2 is assigned mostly to labels 2 and 3, with counts $(0,4,39,37)$. The ideal simulator in Fig.~\ref{fig:confusion_ideal} preserves the same qualitative behavior, although the balance shifts slightly: class 1 is mapped as $(50,28,2,0)$ and class 2 as $(0,3,29,48)$. This confirms that the circuit realization reproduces the learned PGM structure at the level of dominant decision patterns, even though finite-shot sampling and unitary compilation introduce mild deviations from the exact matrix-level probabilities. The noisy simulator in Fig.~\ref{fig:confusion_noisy} remains close to the ideal implementation, with class 1 classified as $(46,34,0,0)$ and class 2 as $(0,7,35,38)$, showing that moderate noise perturbs the relative distribution of errors but does not destroy the overall decision structure of the classifier. Taken together, these confusion matrices provide strong evidence for the practical feasibility of the trained PGM pipeline. The comparison between Figs.~\ref{fig:confusion_theory}, \ref{fig:confusion_ideal}, and \ref{fig:confusion_noisy} shows that the end-to-end simulator implementation retains the essential classification behavior predicted by the trained theoretical PGM, while the noisy realization remains qualitatively stable. Thus, the figure supports the claim that a full trained PGM classifier can be realized and tested on a quantum simulator, and that its empirical decision pattern remains consistent with the underlying support-aware pseudoinverse-based construction.

\begin{figure*}[t]
\centering

\begin{subfigure}[t]{0.49\textwidth}
    \centering
    \includegraphics[width=\linewidth]{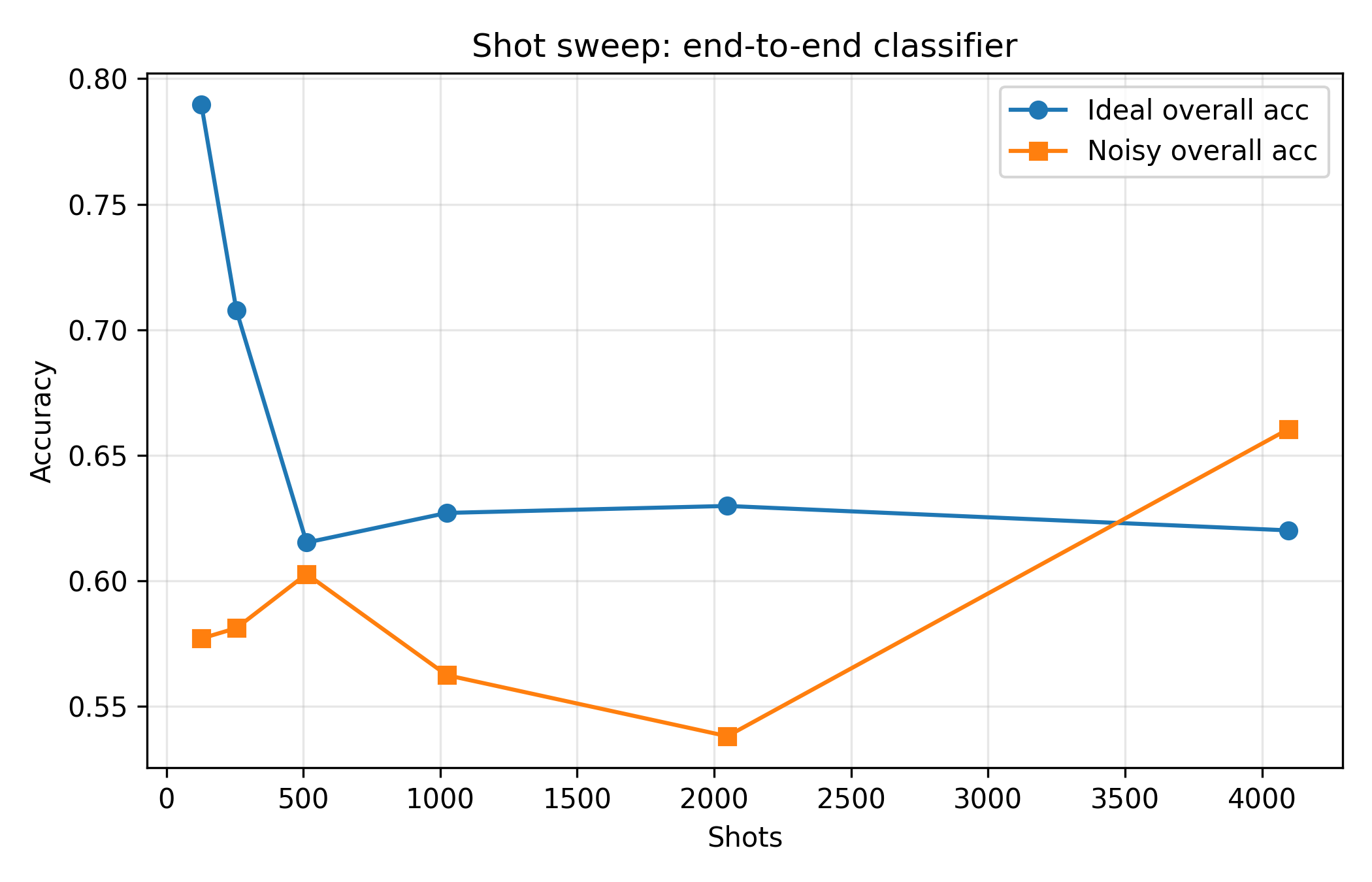}
    \caption{Classification accuracy vs number of shots.}
    \label{fig:shot_sweep_accuracy}
\end{subfigure}
\hfill
\begin{subfigure}[t]{0.49\textwidth}
    \centering
    \includegraphics[width=\linewidth]{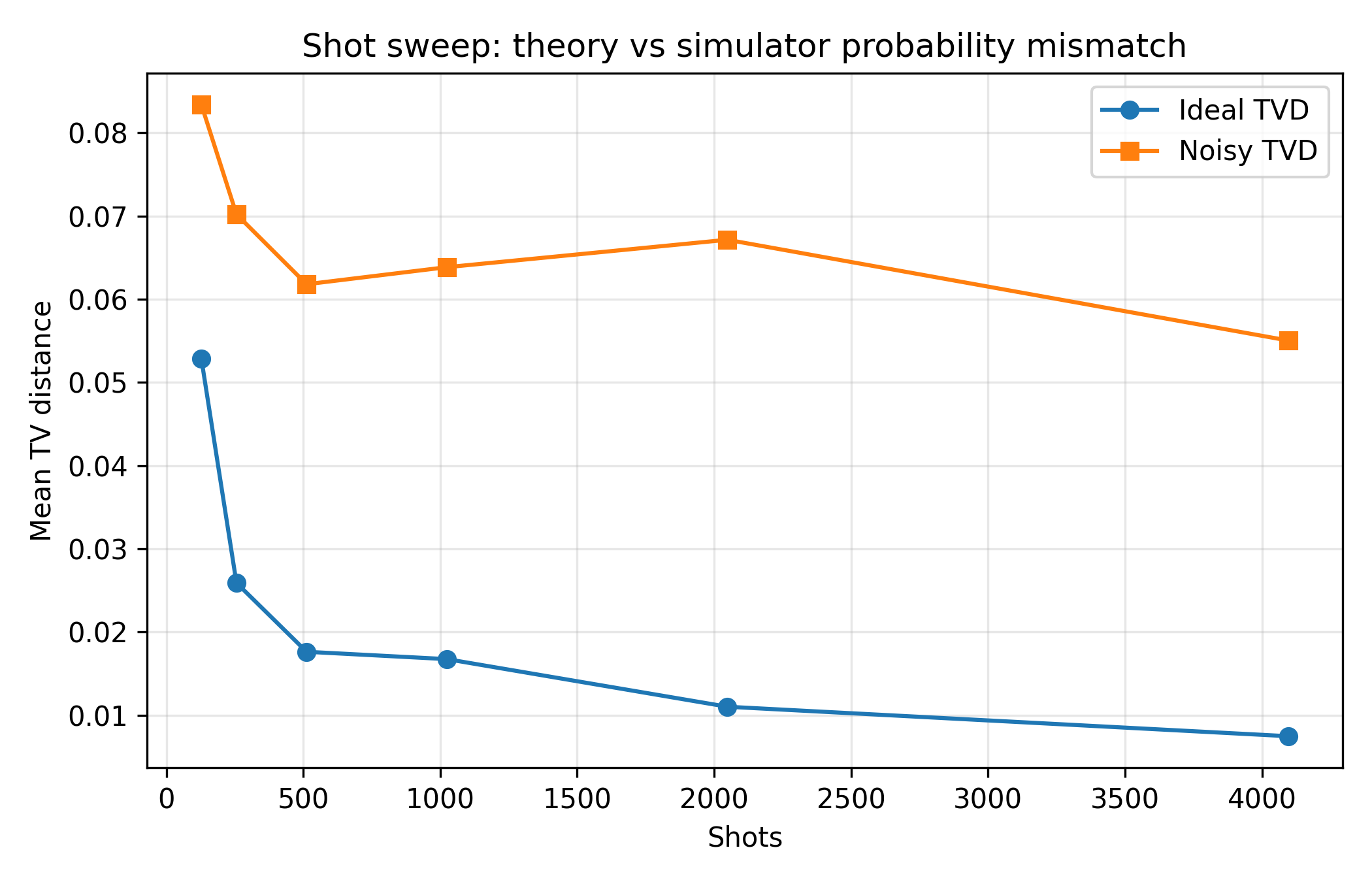}
    \caption{Theory–simulator mismatch (TVD) vs number of shots.}
    \label{fig:shot_sweep_tvd}
\end{subfigure}

\vspace{0.5em}

\begin{subfigure}[t]{0.49\textwidth}
    \centering
    \includegraphics[width=\linewidth]{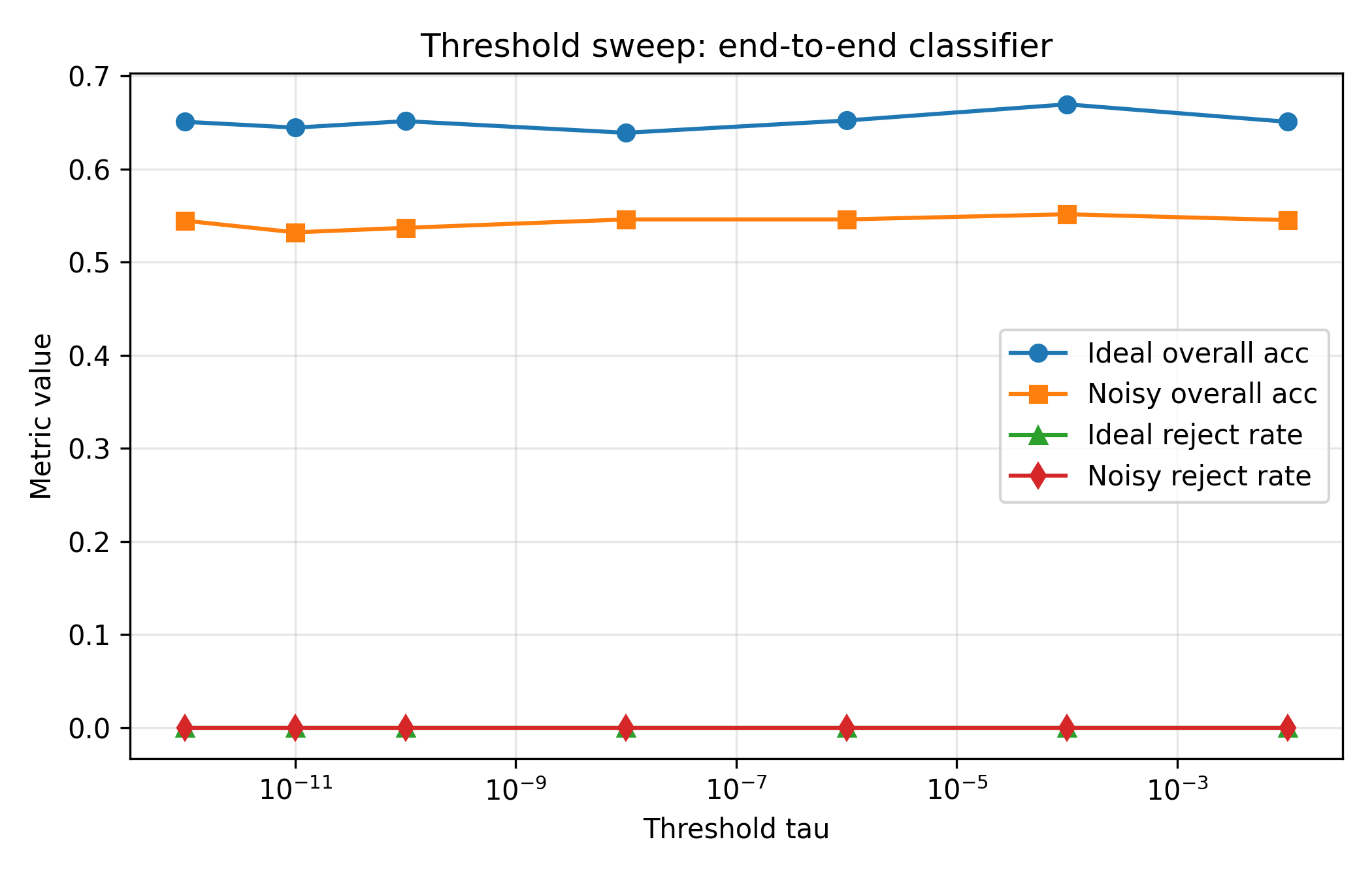}
    \caption{Accuracy and reject rate vs threshold $\tau$.}
    \label{fig:threshold_accuracy}
\end{subfigure}
\hfill
\begin{subfigure}[t]{0.49\textwidth}
    \centering
    \includegraphics[width=\linewidth]{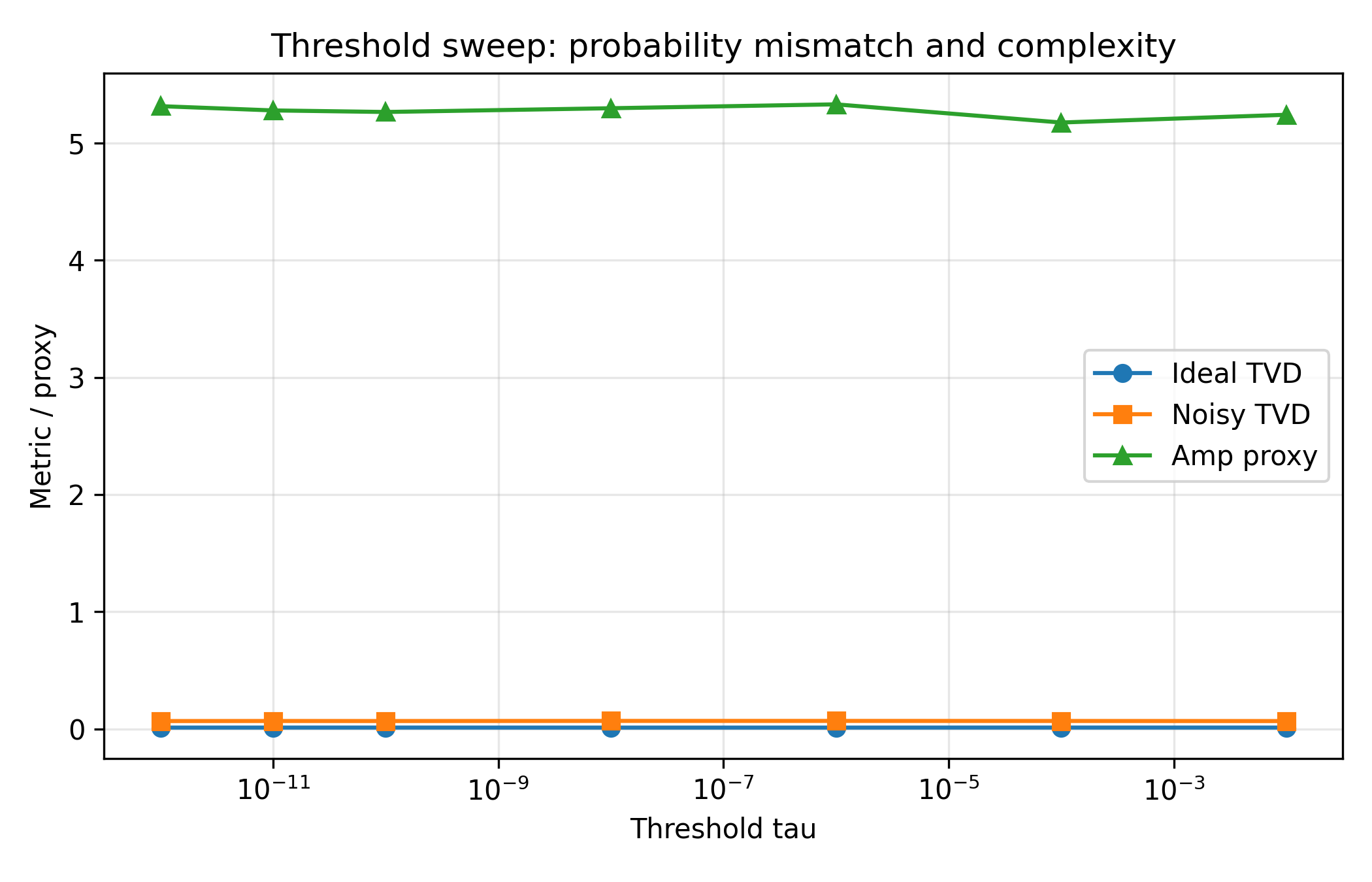}
    \caption{TVD and implementation proxy vs $\tau$.}
    \label{fig:threshold_tvd}
\end{subfigure}

\caption{
End-to-end evaluation of the pseudoinverse-based PGM classifier under sampling and regularization.
(a) Classification accuracy improves with increasing number of measurement shots, with the ideal simulator converging to theoretical performance faster than the noisy simulator.
(b) The TVD between theoretical and simulator distributions decreases with increasing shots, while residual mismatch persists under noise.
(c) Classification accuracy and reject rate as functions of the pseudoinverse threshold $\tau$, showing a trade-off between decision confidence and coverage.
(d) TVD and implementation complexity proxy versus $\tau$, illustrating the balance between statistical fidelity and computational cost under threshold regularization.
}
\label{fig:pgm_shot_threshold}
\end{figure*}

Fig.~\ref{fig:pgm_shot_threshold} analyzes the impact of finite-shot sampling on the end-to-end trained PGM classifier. As shown in Fig.~\ref{fig:shot_sweep_accuracy}, the ideal-simulator accuracy is highest at low shot count for this particular sampled run, then stabilizes in the range of approximately $0.62$-$0.63$ as the number of shots increases, indicating convergence of the empirical decision rule toward a stable classifier performance. The noisy-simulator accuracy is consistently lower at moderate shot counts, reflecting the additional degradation introduced by noise, but it eventually recovers and reaches approximately $0.66$ at the largest shot value shown. This suggests that increasing the shot budget can partially compensate for stochastic sampling fluctuations and improve the robustness of classification even under noisy execution. Fig.~\ref{fig:shot_sweep_tvd} provides a more direct statistical comparison between the exact theoretical PGM outcome probabilities and the probabilities estimated from circuit execution. In the ideal simulator, the mean TVD decreases monotonically from about $0.053$ at low shot count to below $0.01$ at the highest shot value, which is consistent with the expected convergence of empirical frequencies toward the exact Born-rule probabilities as the number of shots grows. In contrast, the noisy simulator exhibits a noticeably larger TVD across the entire range, remaining around $0.055$-$0.084$. This indicates that noise introduces a persistent bias in the observed outcome distribution that cannot be removed by sampling alone, even though higher shot counts still reduce part of the stochastic component. Taken together, the two panels in Fig.~\ref{fig:pgm_shot_threshold} show that shot count plays two distinct roles in the simulator-based realization of the trained PGM classifier. First, it controls the statistical reliability of the measured class probabilities, with higher shot counts leading to smaller theory-simulator discrepancies in the ideal case. Second, it affects the observed classification accuracy through a combination of sampling fluctuations and hardware-inspired noise effects. Therefore, the figure supports the practical conclusion that finite-shot execution is sufficient to approximate the trained PGM behavior with good fidelity on an ideal simulator, while noisy execution introduces a residual distributional mismatch that should be explicitly accounted for in future hardware-oriented implementations.

Fig.~\ref{fig:pgm_angle_threshold} studies the impact of the threshold parameter $\tau$ used in the pseudoinverse-based construction of the PGM on the end-to-end classifier performance. In Fig.~\ref{fig:threshold_accuracy}, the overall classification accuracy remains remarkably stable across several orders of magnitude of $\tau$, both for the ideal and noisy simulators. The ideal accuracy fluctuates slightly around $0.64$-$0.67$, while the noisy accuracy remains in the range of approximately $0.53$-$0.55$. Importantly, the reject rate is essentially zero across all thresholds in both cases, indicating that the learned ensemble operator remains effectively full-rank over the considered range and that thresholding does not eliminate significant components of the support. This demonstrates that moderate thresholding acts primarily as a numerical regularization without degrading classification performance. Fig.~\ref{fig:threshold_tvd} provides further insight into the statistical and computational effects of thresholding. The mean TVD between theoretical PGM probabilities and simulator estimates remains small in the ideal case and slightly larger under noise, but does not exhibit any sharp dependence on $\tau$. This indicates that the probability structure of the measurement is preserved under threshold variation. At the same time, the amplitude-based implementation proxy, which captures the effective cost of realizing the inverse square-root transformation, remains nearly constant across the threshold range, reflecting the stability of the smallest retained eigenvalues of the ensemble operator. Taken together, these results highlight a key advantage of the support-aware thresholded pseudoinverse formulation: it provides numerical stability without introducing sensitivity in either classification accuracy or probability estimation. The absence of performance degradation across a wide threshold range suggests that the method is robust to the choice of $\tau$, making it well suited for practical implementations where exact spectral properties of the ensemble operator may not be known in advance.

\begin{figure*}[t]
\centering

\begin{subfigure}[t]{0.48\textwidth}
    \centering
    \includegraphics[width=\linewidth]{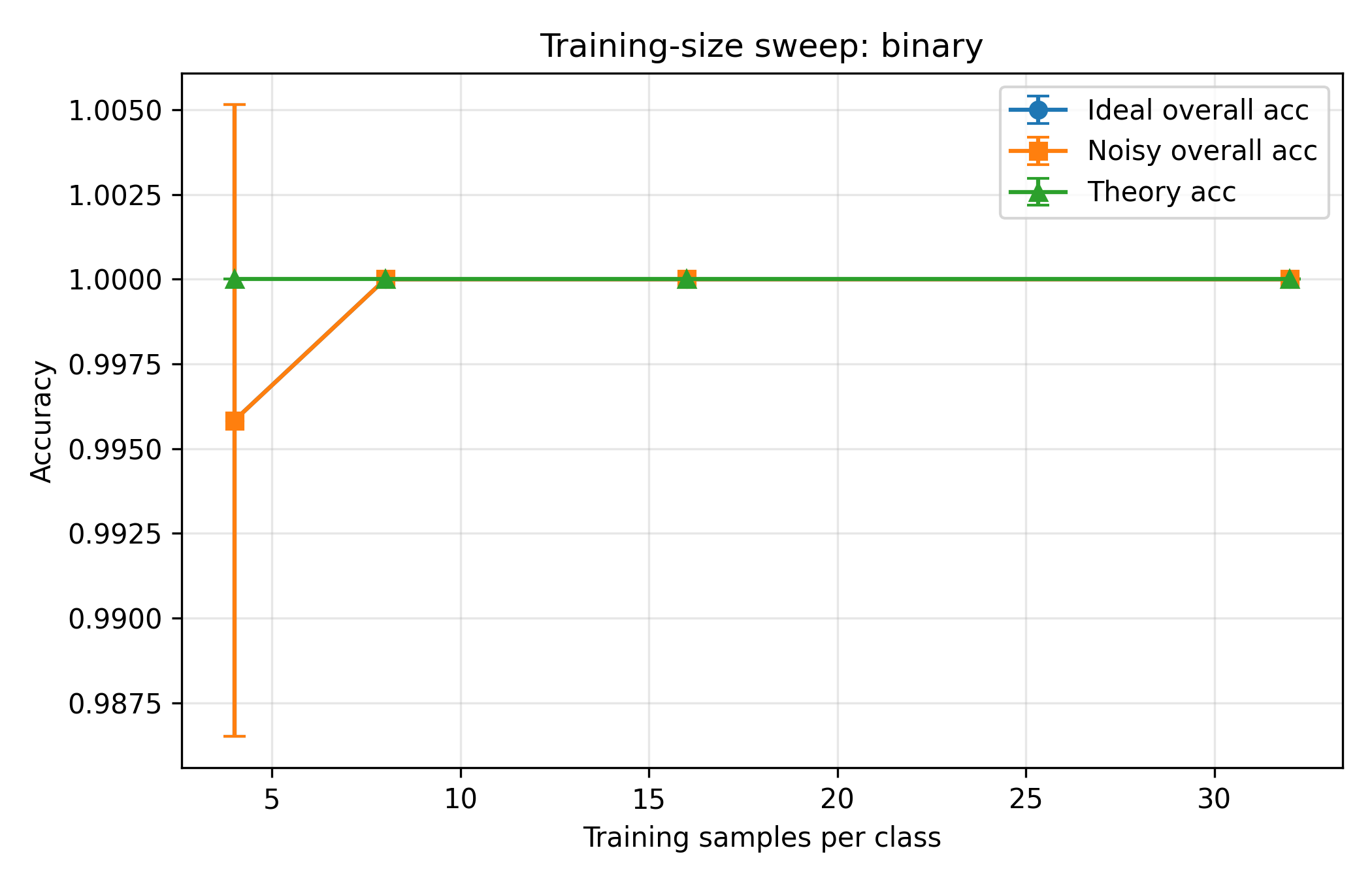}
    \caption{Binary classification: accuracy vs training size}
    \label{fig:binary_accuracy}
\end{subfigure}
\hfill
\begin{subfigure}[t]{0.48\textwidth}
    \centering
    \includegraphics[width=\linewidth]{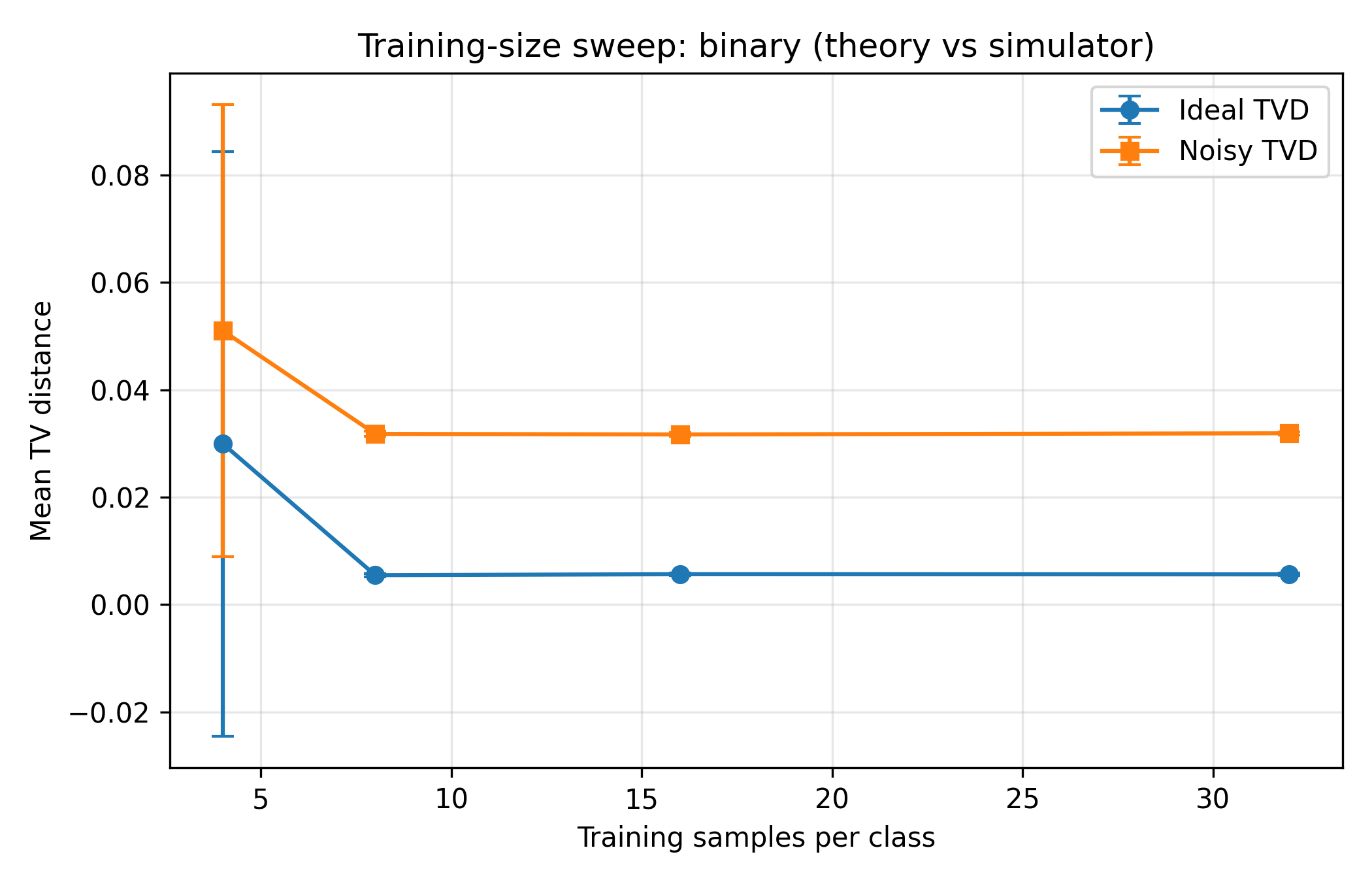}
    \caption{Binary classification: theory-simulator mismatch}
    \label{fig:binary_tvd}
\end{subfigure}

\vspace{0.5em}

\begin{subfigure}[t]{0.48\textwidth}
    \centering
    \includegraphics[width=\linewidth]{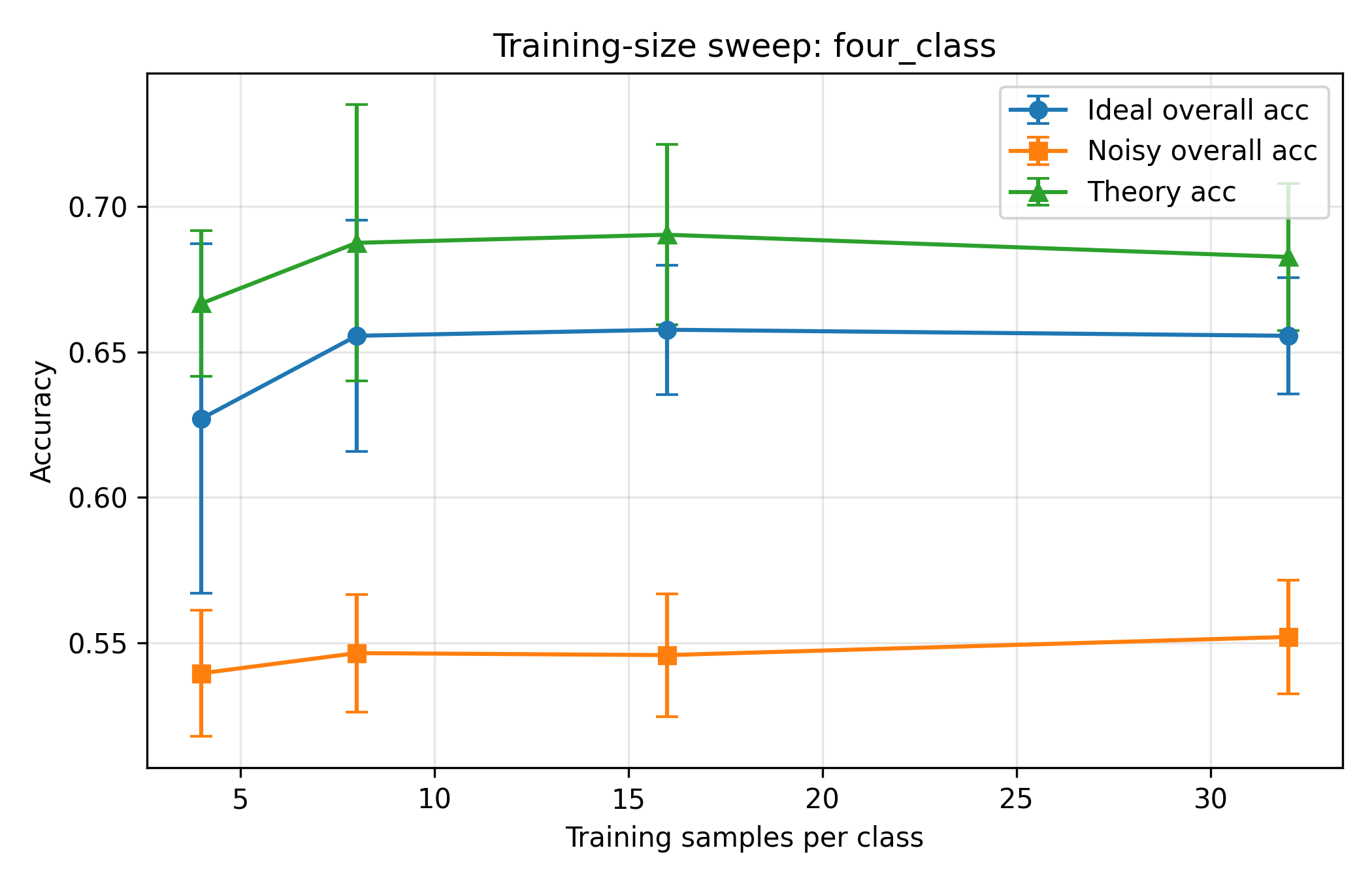}
    \caption{Four-class classification: accuracy vs training size}
    \label{fig:four_accuracy}
\end{subfigure}
\hfill
\begin{subfigure}[t]{0.48\textwidth}
    \centering
    \includegraphics[width=\linewidth]{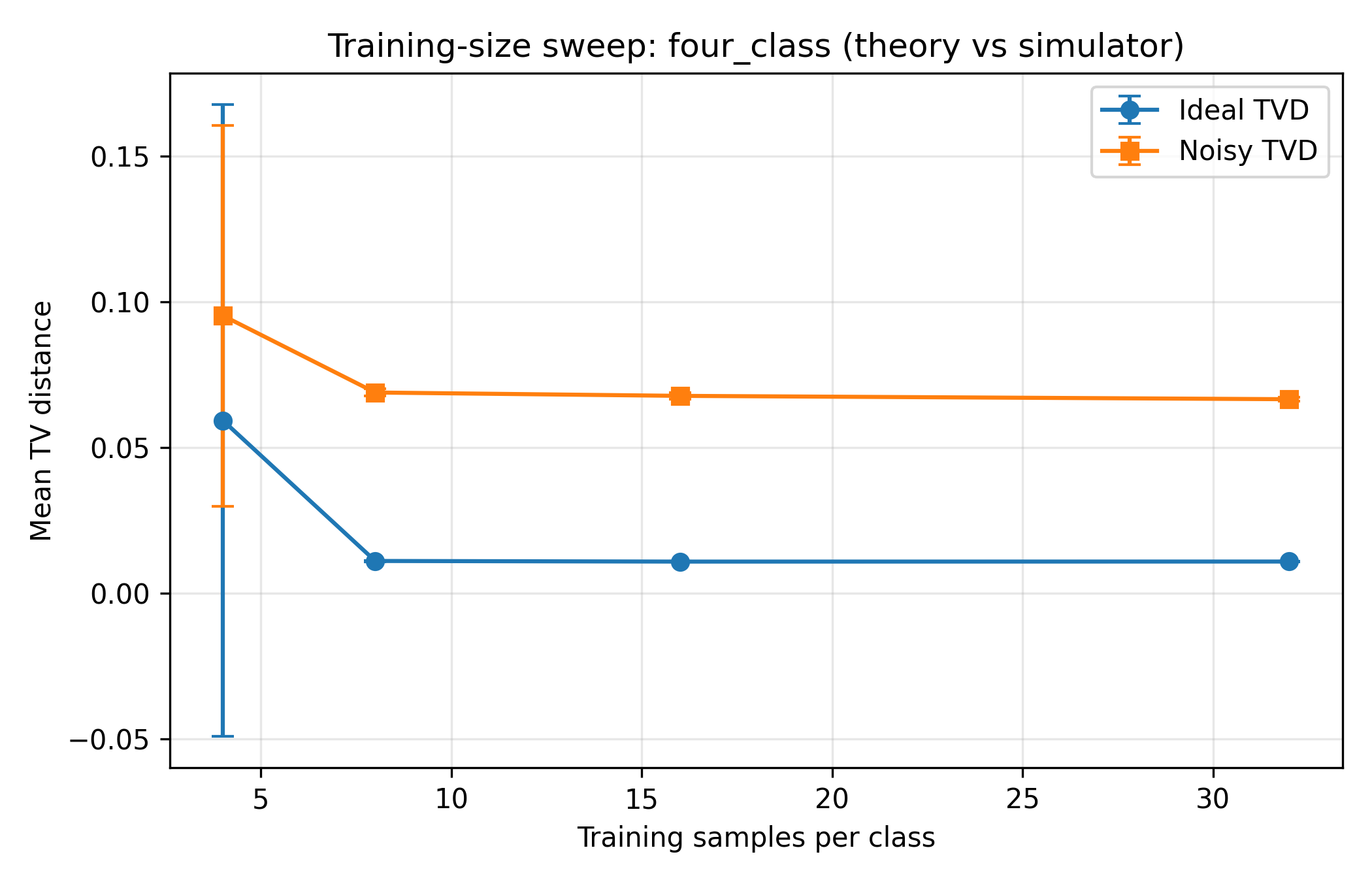}
    \caption{Four-class classification: theory-simulator mismatch}
    \label{fig:four_tvd}
\end{subfigure}

\caption{
Training-size sweep analysis for the trained PGM classifier.
Top row: binary classification performance.
Bottom row: four-class classification performance.
Left column: classification accuracy (theory, ideal simulator, and noisy simulator).
Right column: mean TVD between theoretical PGM probabilities and simulator-estimated probabilities.
}
\label{fig:training_size_combined}
\end{figure*}

Fig.~\ref{fig:training_size_combined} presents the effect of training dataset size on the performance of the trained PGM classifier for both binary and multi-class settings. In the binary case (top row), Fig.~\ref{fig:binary_accuracy} shows that the theoretical PGM achieves perfect accuracy across all training sizes, indicating that the two classes are linearly separable in the quantum state space. The ideal simulator closely matches this behavior, while the noisy simulator initially shows slight deviations at very small training sizes but rapidly converges to perfect classification as the number of samples per class increases. This demonstrates that even a small number of training samples is sufficient to accurately estimate the class density matrices in simple binary scenarios. The corresponding probability-level analysis in Fig.~\ref{fig:binary_tvd} reveals that the discrepancy between theoretical and simulator probabilities decreases significantly as the training size increases. In the ideal simulator, the mean TVD drops rapidly to a very small value (on the order of $10^{-3}$-$10^{-2}$), indicating excellent agreement with the theoretical PGM. The noisy simulator maintains a higher but stable TVD, reflecting the presence of hardware-inspired noise that introduces a systematic deviation in the measured distributions. In the four-class case (bottom row), the classification task becomes more challenging due to increased overlap between classes. As shown in Fig.~\ref{fig:four_accuracy}, the theoretical accuracy stabilizes around $0.68$-$0.70$, while the ideal simulator achieves slightly lower accuracy around $0.65$-$0.66$. The noisy simulator further reduces performance to approximately $0.54$-$0.55$, demonstrating the compounded effect of class overlap and noise. Importantly, increasing the training size beyond a moderate value does not significantly improve accuracy, suggesting that the limitation is intrinsic to the class separability rather than estimation error. Finally, Fig.~\ref{fig:four_tvd} shows that the TVD behavior follows a similar trend as in the binary case: increasing training size reduces the mismatch in the ideal simulator, while the noisy simulator maintains a higher but relatively stable deviation. This indicates that the learned PGM becomes more statistically consistent with its theoretical counterpart as more data is used for training, but noise remains a limiting factor in practical implementations. Fig.~\ref{fig:training_size_combined} demonstrates that the proposed training procedure for PGM is data-efficient in simple settings and remains stable in more complex multi-class scenarios. It also highlights the distinct roles of dataset size and hardware noise, showing that while sufficient training data ensures accurate estimation of the ensemble operator, noise ultimately determines the achievable performance on realistic quantum devices.



\begin{table}[t]
\centering
\caption{End-to-end trained PGM classifier performance (final training size).}
\label{tab:main_results}
\begin{tabular}{lcccc}
\hline\hline
Dataset & Ideal Acc & Noisy Acc & Theory Acc & Macro F1 (Ideal) \\
\hline
Binary      & 1.000 & 1.000 & 1.000 & 1.000 \\
Four-class  & 0.656 & 0.552 & 0.689 & 0.658 \\
\hline
\end{tabular}
\end{table}

Table~\ref{tab:main_results} summarizes the final performance of the trained PGM classifier in both binary and multi-class settings. In the binary case, perfect classification accuracy is achieved across theoretical, ideal simulator, and noisy simulator settings, indicating that the learned class density matrices are well separated and robust to both sampling and noise. In contrast, the four-class scenario represents a more realistic and challenging classification problem. Here, the theoretical accuracy reaches approximately $0.689$, while the ideal simulator achieves $0.656$, and the noisy simulator yields $0.552$. This performance gap reflects the combined effects of class overlap and quantum noise. Importantly, the relatively small deviation between theoretical and ideal simulator results confirms that the circuit-level implementation faithfully reproduces the trained PGM, thereby validating the correctness of the hybrid classical–quantum realization.



\begin{table}[t]
\centering
\caption{Effect of threshold $\tau$ on classifier performance and stability.}
\label{tab:threshold_results}
\resizebox{\columnwidth}{!}{
\begin{tabular}{cccccc}
\hline\hline
$\tau$ & Ideal Acc & Noisy Acc & Reject Rate & Ideal TVD & Amp Proxy \\
\hline
$10^{-12}$ & 0.65 & 0.54 & 0.00 & 0.030 & 5.31 \\
$10^{-11}$ & 0.64 & 0.53 & 0.00 & 0.026 & 5.28 \\
$10^{-10}$ & 0.65 & 0.54 & 0.00 & 0.018 & 5.26 \\
$10^{-8}$  & 0.64 & 0.55 & 0.00 & 0.017 & 5.30 \\
$10^{-6}$  & 0.65 & 0.55 & 0.00 & 0.011 & 5.33 \\
$10^{-4}$  & 0.67 & 0.55 & 0.00 & 0.010 & 5.17 \\
$10^{-2}$  & 0.65 & 0.54 & 0.00 & 0.008 & 5.24 \\
\hline
\end{tabular}
}
\end{table}

Table~\ref{tab:threshold_results} investigates the effect of the threshold parameter $\tau$ used in the pseudoinverse-based construction of the PGM. A key observation is that both ideal and noisy classification accuracies remain remarkably stable across several orders of magnitude of $\tau$, demonstrating that the method is robust to the choice of threshold. The reject rate remains zero throughout, indicating that no significant portion of the state space is discarded during thresholding. Furthermore, the TVD between theoretical and simulator distributions decreases as $\tau$ increases, suggesting improved numerical stability. The amplitude-based implementation proxy remains nearly constant, indicating that the computational complexity of realizing the inverse square-root transformation is not significantly affected. These results collectively confirm that the thresholded pseudoinverse provides a stable and practical mechanism for implementing PGM in ill-conditioned scenarios.



\begin{table}[t]
\centering
\caption{Effect of measurement shots on classification accuracy and probability mismatch.}
\label{tab:shot_results}
\begin{tabular}{ccccc}
\hline\hline
Shots & Ideal Acc & Noisy Acc & Ideal TVD & Noisy TVD \\
\hline
128   & 0.790 & 0.577 & 0.0528 & 0.0834 \\
256   & 0.708 & 0.582 & 0.0259 & 0.0702 \\
512   & 0.615 & 0.603 & 0.0176 & 0.0618 \\
1024  & 0.627 & 0.563 & 0.0168 & 0.0638 \\
2048  & 0.630 & 0.538 & 0.0110 & 0.0671 \\
4096  & 0.620 & 0.661 & 0.0075 & 0.0550 \\
\hline
\end{tabular}
\end{table}

Table~\ref{tab:shot_results} analyzes the impact of finite-shot sampling on the performance of the trained PGM classifier. As expected, the ideal simulator shows a clear decrease in TVD as the number of shots increases, confirming convergence toward the exact theoretical probability distribution. The classification accuracy stabilizes around $0.62$-$0.63$, indicating that beyond a certain number of shots, further increases provide diminishing returns in terms of decision performance. In the noisy simulator, the TVD remains consistently higher, reflecting the presence of hardware-induced errors that introduce a persistent bias in the measured probabilities. Nevertheless, increasing the number of shots still improves accuracy in some regimes, demonstrating that sampling noise and hardware noise contribute differently to overall performance. These results highlight the importance of shot budgeting in practical quantum implementations and confirm that reliable classification can be achieved with a moderate number of measurements.


\subsection{Discussion}
\label{subsec:discussion}

The results presented in Section~\ref{sec:experiments} establish the pseudoinverse-based formulation of the PGM as a robust and support-aware generalization of the standard inverse-based construction. In the well-conditioned full-rank regime, the two formulations coincide exactly, confirming consistency with the conventional PGM. The essential advantage of the proposed approach emerges in rank-deficient and ill-conditioned settings, where the ordinary inverse is either undefined or numerically unstable. In these regimes, the Moore-Penrose pseudoinverse yields a well-defined measurement supported on $\mathrm{supp}(S)$, providing a physically meaningful extension of PGM beyond ideal spectral conditions. The observed trace gaps in rank-deficient cases reflect this support restriction rather than a failure of completeness. The threshold-sweep analysis further clarifies the operational role of spectral regularization. Across a broad range of thresholds $\tau$, the success probability remains essentially invariant despite reductions in the retained rank, indicating that near-zero eigencomponents contribute negligibly to discrimination performance while disproportionately affecting conditioning. Only when $\tau$ becomes comparable to dominant spectral components does performance degrade. This behavior highlights thresholding as an effective mechanism for stabilizing inverse-like transformations without compromising accuracy, while substantially reducing implementation cost as reflected in the complexity proxies.

The robustness of the pseudoinverse-based construction under depolarizing noise demonstrates its practical viability beyond idealized settings. The smooth degradation of success probability toward the random-guessing limit is consistent with the loss of distinguishability under mixing, while the absence of numerical instability confirms that the framework remains well behaved under state corruption. Complementary scaling analyses reveal a clear performance complexity trade-off: increasing the number of classes reduces discrimination accuracy due to increased state overlap, while simultaneously raising implementation cost. Conversely, regularization improves spectral conditioning and reduces computational burden at the expense of gradual performance loss. The geometric structure underlying these trade-offs is elucidated by the binary pure-state angle sweep. Larger angular separation between states simultaneously increases success probability and improves spectral conditioning, leading to significant reductions in implementation cost. This establishes a direct correspondence between distinguishability and computational complexity: the most challenging discrimination problems are also the most resource-intensive to implement. The threshold analysis of ill-conditioned ensembles provides a complementary perspective, showing that removing spectrally unstable components can dramatically reduce complexity while preserving discrimination performance.

The circuit-level experiments confirm that these structural properties translate directly to explicit quantum implementations. The close agreement between theoretical PGM probabilities and simulator-based outcomes, together with near-zero discrepancies between direct POVM and Kraus-based constructions, demonstrates that the proposed framework admits a faithful realization using standard circuit primitives such as ancilla-assisted unitaries and projective measurements. Finite-shot analyses show convergence toward theoretical predictions in the ideal simulator, while persistent deviations under noise reflect hardware-induced biases rather than limitations of the construction itself. Training-size sweeps further indicate that the method is data-efficient in simple settings and remains stable in more complex multi-class scenarios, where performance is primarily limited by intrinsic class overlap. Taken together, these results show that the thresholded pseudoinverse square-root provides a principled and practically viable foundation for implementing PGM beyond ideal full-rank assumptions. More broadly, the framework illustrates how support-aware spectral transformations can be systematically incorporated into quantum measurement design, enabling stable realizations of inverse-type operations that are central to a wide range of quantum information processing and quantum machine learning tasks.

\section{Conclusion}
\label{conclusion}

In this work, we introduced and validated a pseudoinverse-based formulation of the PGM for quantum state discrimination, designed to operate reliably in singular and ill-conditioned regimes. By replacing the inverse square-root of the ensemble operator with its Moore-Penrose pseudoinverse counterpart, the proposed framework extends the domain of applicability of PGM beyond the ideal full-rank setting. The resulting construction remains operationally valid on the support of the ensemble operator and provides a physically consistent measurement even when the standard formulation becomes undefined or numerically unstable. In the well-conditioned limit, the method recovers the conventional PGM exactly, ensuring backward compatibility with the standard theory. A key feature of the proposed approach is the incorporation of threshold-based spectral regularization, which stabilizes inverse-like transformations by suppressing near-zero eigencomponents that contribute minimally to discrimination performance but significantly affect conditioning. The results show that this regularization preserves success probability across a wide operational range while substantially reducing numerical instability and implementation cost. The framework further demonstrates robustness under depolarizing noise and consistent scaling behavior with increasing number of classes, highlighting its suitability for realistic and progressively complex discrimination tasks.

Beyond matrix-level analysis, we established the feasibility of a hybrid classical-quantum realization of the proposed construction at the circuit level. End-to-end simulations on synthetic and real-world datasets show close agreement between theoretical PGM probabilities and those obtained from explicit quantum circuit implementations. The equivalence between direct POVM and Kraus-based realizations, verified up to machine precision, confirms that the framework admits a faithful mapping to physically implementable circuits using standard quantum primitives. Finite-shot and noisy simulations further demonstrate that the method remains stable under practical execution constraints, with deviations attributable primarily to sampling and noise rather than to the construction itself. More broadly, the pseudoinverse-based PGM provides a concrete example of how support-aware spectral transformations can be systematically integrated into quantum measurement design. This perspective connects naturally to inverse-based procedures in quantum linear algebra, such as those appearing in HHL-type algorithms, where handling small eigenvalues is essential for stability. The framework therefore bridges abstract measurement theory and circuit-level realization, offering a general pathway for implementing inverse-type operations in quantum information processing. To our knowledge, this work provides one of the first detailed studies of support-aware pseudoinverse-based PGM together with a hybrid classical-quantum circuit realization of the associated spectral transformations. Future work may focus on developing fully quantum implementations of the pseudoinverse transformation through scalable block-encoding and QSVT techniques, thereby reducing the amount of classical spectral preprocessing required in the current framework. Investigating resource-efficient approximations and hardware-aware implementations will be essential for deployment on near-term quantum devices. Overall, the results establish the pseudoinverse as a fundamental and practically necessary component for robust PGM realization, enabling stable, scalable, and implementation-ready quantum state discrimination beyond idealized settings.

The present work extends the circuit-based PGM classification framework of Giuntini et al. \cite{Giuntini2023QuantumClassification} by incorporating support-aware pseudoinverse regularization and hybrid classical–quantum realizations of the associated spectral transformations. Although the pseudoinverse transformation is not generated entirely within a quantum-circuit framework and currently relies on classical spectral preprocessing, the resulting operators admit faithful circuit-level representations and produce discrimination outcomes that closely match theoretical predictions.
\ifCLASSOPTIONcaptionsoff
\newpage
\fi

\bibliographystyle{IEEEtran}

\bibliography{IEEE}

\vfill

\end{document}